\documentclass[10pt,journal,compsoc]{IEEEtran}
\usepackage{amsmath,amssymb}
\usepackage{algorithm}
\usepackage{algpseudocode}
\usepackage{amsthm}
\usepackage{booktabs}
\usepackage{graphicx}
\usepackage{multirow}
\usepackage{url}
\usepackage{xcolor}
\newcommand{\vrmaas}{VRMaaS}
\newcommand{\rev}[1]{#1}
\newcommand{\brev}[1]{#1}
\definecolor{revisionyellow}{HTML}{B8860B}
\newcommand{\yrev}[1]{#1}
\usepackage{hyperref}
\hypersetup{pdftitle={VRMaaS: Verifiable Reward-Model-as-a-Service for Trustworthy LLM Alignment},pdfauthor={Guofu Liao, Hailang Xi, Taotao Wang, Shengli Zhang, Rongcheng Tu, Hao Liao}}
\newtheorem{definition}{Definition}
\newtheorem{lemma}{Lemma}
\newtheorem{theorem}{Theorem}

\begin{document}

\title{\vrmaas: Verifiable Reward-Model-as-a-Service for Trustworthy LLM Alignment}

\author{\rev{Guofu~Liao, Hailang~Xi, Taotao~Wang, Shengli~Zhang, Rongcheng~Tu, and Hao~Liao}%
\thanks{\rev{Guofu Liao, Hailang Xi, Taotao Wang, and Shengli Zhang are with the College of Electronics and Information Engineering, Shenzhen University, China (e-mail: liaoguofu2022@email.szu.edu.cn; 2400042055@mails.szu.edu.cn; ttwang@szu.edu.cn; zsl@szu.edu.cn). Guofu Liao and Hailang Xi contributed equally to this work. Corresponding author: Taotao Wang.}}%
\thanks{\rev{Rongcheng Tu is with the College of Computing and Data Science, Nanyang Technological University, Singapore (e-mail: rongcheng.tu@ntu.edu.sg).}}%
\thanks{\rev{Hao Liao is with the College of Computer Science and Software Engineering, Shenzhen University, China (e-mail: haoliao@szu.edu.cn).}}%
}

\markboth{VRMaaS --- arXiv Preprint, November 2026}%
{\vrmaas: Verifiable Reward-Model-as-a-Service for Trustworthy LLM Alignment}

\IEEEtitleabstractindextext{%
\begin{abstract}
Reward models (RMs) are increasingly deployed as service components in LLM alignment pipelines, where payment, audit, moderation, and governance workflows act on the scores they return. An inference proof alone does not show whether a score belongs to the intended request, model checkpoint, numerical configuration, and verifier: in a controlled comparison that isolates the accepted statement from prover scale, a generic inference proof wrapped for on-chain checking remains valid after six classes of such service-field substitution. We formulate \emph{verifiable reward-model-as-a-service} (\vrmaas) as a service-output integrity problem and present a framework in which each returned reward becomes a proof-carrying artifact. Its off-chain acceptance rule binds request identity, committed weights, verifier and configuration identifiers, proof-task order, and the digest of the exported package, so auditors can check a reward without rerunning the private RM or exposing inputs, checkpoints, or hidden activations; an optional on-chain zero-knowledge wrapper gives contracts a partial attestation of selected consistency checks and supplied service fields. The framework decomposes Transformer RM serving into arithmetic and lookup-backed proof tasks, aggregates them into a composite proof, and links hidden operands to the committed weights. A GPU prototype on 1B, 3B, and 8B reward models over 128--1024 tokens preserves 93.5--99\% pairwise agreement with the baseline RM and rejects every tested service-field substitution in complete artifacts. Using exact integer witnesses and request-level aggregation of weight-binding openings, it generates complete artifacts in 49.8--206.1 seconds across the evaluated settings; auditors verify retained artifacts in 1.94--5.11 seconds, and partial public attestation costs about 217K gas. These costs suit delayed audit, settlement, and governance workflows.
\end{abstract}

\begin{IEEEkeywords}
\rev{Services computing, reward-model serving, AI-service output integrity, proof-carrying artifacts, verifiable computation, public attestation.}
\end{IEEEkeywords}}

\maketitle

\IEEEdisplaynontitleabstractindextext
\IEEEpeerreviewmaketitle

\section{Introduction}
\IEEEPARstart{L}{arge} language models (LLMs) have moved into service infrastructure for ranking, moderation, response selection, and downstream automation~\rev{\cite{achiam2023gpt,liang2025widespread}}. Many LLM preference-alignment pipelines rely on reward models (RMs), which score prompt-response pairs~\rev{\cite{liu2025skywork}} and support reinforcement learning from human feedback (RLHF)~\rev{\cite{stiennon2020learning,ouyang2022training}}. \rev{In deployment, the RM becomes an AI-service result producer whose scores are consumed by requesters, auditors, payment logic, moderation workflows, governance systems, and service contracts.}

\rev{This role creates a service-output integrity problem: a conventional RM API returns a score, not an accepted statement binding it to the intended request, checkpoint, numerical configuration, and verifier logic~\cite{zhao2021veriml,huang2022zkmlaas}.} \rev{A provider may substitute a cheaper model, replay a stale result, alter configuration, or graft a valid-looking output onto the wrong context.} We call this setting \emph{verifiable reward-model-as-a-service} (\vrmaas): the service returns a reward plus a request-bound artifact that is verifiable across service boundaries while \brev{keeping private inputs, checkpoints, and hidden activations outside the public statement and making off-chain audit explicit}.

From a services-computing perspective, \vrmaas{} concerns the lifecycle of an AI-service output rather than the RM computation alone~\rev{\cite{huhns2005service,papazoglou2007service}}. The reward artifact acts as a service contract across invocation, delivery, audit, and settlement: request binding supports accountability, model/configuration/verifier binding supports service-version management, off-chain verification supports third-party audit, and optional public attestation supports settlement among parties without shared infrastructure trust. \rev{Thus, the problem is not simply whether a neural-network inference can be proved, but whether the accepted reward object remains portable, auditable, and contract-consumable after it crosses organizational boundaries.}

Existing work addresses related pieces but not this service abstraction. Reward-model evaluation and verifiable-reward methods ask whether a reward signal or task outcome is meaningful~\cite{malik2025rewardbench2advancingreward,liu2025trust}, while verifiable ML and ZKML systems prove neural-network computation on untrusted infrastructure~\cite{ghodsi2017safetynets,zhao2021veriml,huang2022zkmlaas,sun2024zkllm}. \rev{A deployed RM service also needs request, model, configuration, verifier, proof-task order, and export semantics bound into the accepted service output. Fields recorded only as provider metadata do not constrain downstream acceptance unless the verification rule binds and checks them. In our runnable GenericProof+Wrapper baseline, which wraps an inference proof for on-chain checking and runs on two small models because it isolates statement semantics rather than prover scale, the proof remains valid after six classes of service-field substitution, whereas the corresponding \vrmaas{} checks reject them. This experiment makes the service-context gap explicit: consumers need an acceptance rule that connects the proved computation to the invocation and deployment they intended to use.}

We present a \vrmaas{} framework for proof-carrying RM serving. The provider runs RM inference, records a deterministic proof trace, decomposes the Transformer RM path into typed arithmetic and lookup-backed proof tasks, aggregates them into a composite proof, and exports an artifact bound to request identity, committed weights, verifier and configuration identifiers, proof-task order, and the export digest. Downstream consumers verify this artifact instead of replaying the private RM computation: auditors apply the complete off-chain acceptance rule, and an optional on-chain wrapper gives contracts a partial attestation of selected consistency checks and supplied service fields to use alongside the auditor decision. The central contribution is this accepted service-contract semantics, which defines the request-, model-, configuration-, verifier-, order-, and export-bound reward object that downstream services may accept.

Under this service-contract view, this paper makes five contributions.
\begin{itemize}
    \item \rev{We formulate \vrmaas{} as a service-output integrity problem for trustworthy AI-service composition, treating the reward score as a contract-facing service result and defining service roles, trust boundaries, request binding, verifier views, and optional public-attestation interfaces.}
    \item \rev{We define accepted statement semantics that make request identity, model commitment, verifier/configuration identifiers, proof-task order, export digest, and public-statement interpretation first-class acceptance fields.}
    \item \rev{We design an RM-specific proof and binding pipeline that decomposes Transformer RM execution into typed proof tasks, aggregates them into a composite proof, and links hidden operands to authenticated model-weight commitments and declared numerical configuration.}
    \item We develop an efficient artifact-generation pipeline combining exact integer witness construction, reusable lookup workspaces, and request-level batching of weight-binding proofs, and evaluate its producer-side cost.
    \item We implement a prototype on 1B, 3B, and 8B reward models over sequence lengths 128--1024 and evaluate fidelity, service-field rejection against a generic baseline, and producer, auditor, and contract costs. Complete artifacts are generated in 49.8--206.1 seconds, verified by auditors in 1.94--5.11 seconds, and attested partially on chain for about 217K gas, which supports delayed audit, settlement, and governance workflows.
\end{itemize}

\section{Verifiable Reward-Model-as-a-Service}
\label{sec:vrmaas}

\begin{figure*}[t]
    \centering
    \includegraphics[width=0.98\textwidth]{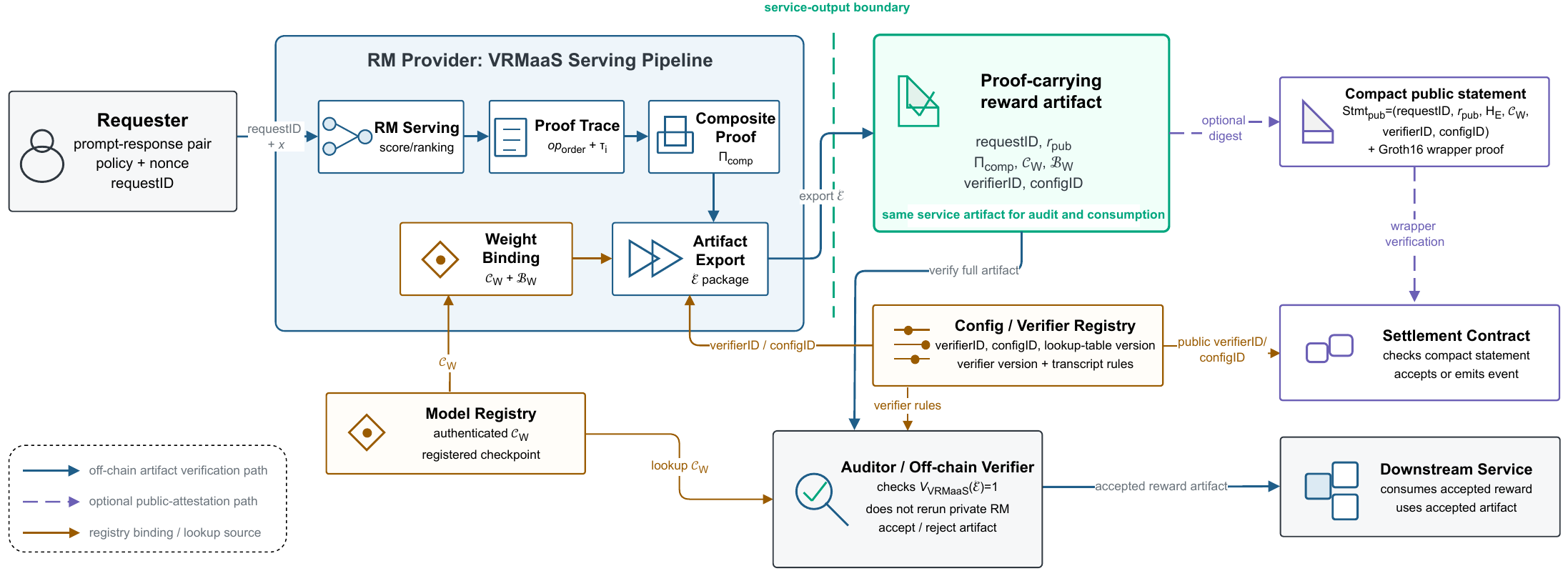}
    \caption{Top-level architecture of \vrmaas. The provider returns a proof-carrying reward artifact bound to the request, model commitment, verifier, and configuration; auditors verify the artifact off chain, and an optional public-attestation layer publishes a compact settlement statement.}
    \label{fig:service-pipeline}
\end{figure*}

This section defines \vrmaas{} as a service abstraction for service-output integrity in reward-model serving. Fig.~\ref{fig:service-pipeline} illustrates the top-level service view of \vrmaas{}.

\subsection{Service Roles}
\vrmaas{} involves four logical roles, shown in Fig.~\ref{fig:service-pipeline}. The \emph{requester} submits a prompt-response pair, comparison workload, \brev{or canonical RM-input package} with a request identity and receives a reward output. The \emph{provider} operates the RM service and \brev{keeps the full checkpoint, private input text, and hidden activations off the public path}. For each accepted invocation, the provider executes the RM, records proof tasks, generates a composite proof, binds proof-relevant weights to a committed model, and exports a request-bound proof-carrying reward artifact. The \emph{auditor} checks whether the returned artifact is valid under the declared service configuration \brev{using the off-chain verification material in the export}. The \emph{public attestation layer}, when used, publishes a compact verification result for multi-party consumption or contract-level settlement.

These roles need not be controlled by separate organizations in every deployment. They define the trust boundaries that a services-computing formulation must expose. The provider wants to avoid \brev{publishing the full proprietary checkpoint}. The requester and auditor want evidence that the score was computed by the declared RM, \brev{with the auditor receiving the opening material needed to validate the exported artifact}. A downstream contract wants a compact public statement rather than a full private inference trace. The artifact produced by the service is therefore the boundary object shared across these roles, allowing them to consume the same reward result without sharing the provider's infrastructure trust.

\subsection{Problem Statement}
The service-level problem is to make an RM score independently checkable after it leaves the provider. The accepted object should carry the context needed by audit, governance, or settlement workflows, not only the numerical reward. Given an untrusted RM service, \vrmaas{} asks \rev{how a requester or auditor can verify that an accepted reward-service output corresponds to the intended RM invocation, committed model weights, declared numerical configuration, verifier logic, and export semantics, while \brev{keeping full private inputs, full checkpoints, and hidden activations out of the public statement and specifying the weight-opening information exposed to off-chain auditors}.}

This question is separate from training a better RM and from proving an arbitrary neural network in isolation. A verifiable RM-service output must be bound to the service identity, request identity, verifier implementation, numerical semantics, and public-export interpretation that downstream consumers rely on. It must remain meaningful after it leaves the original API call and is interpreted by auditors or contract-facing services.

\subsection{Service Scenario and Design Goals}
\rev{Consider a marketplace where workers submit LLM responses, an RM provider scores them, and a settlement service pays workers accordingly. These parties belong to different organizations. A score alone cannot resolve a payment dispute: it gives the settlement service no evidence against a cheaper checkpoint, coarser quantization, or another request's result. We derive five design goals.}
\begin{itemize}
    \item[G1] \emph{Request binding}: an accepted reward names exactly one service job and cannot be replayed or grafted onto another.
    \item[G2] \emph{Model binding}: the reward is computed with a registered checkpoint, without publishing the checkpoint.
    \item[G3] \emph{Configuration accountability}: numerical semantics and verifier logic are part of what is accepted, so they can be versioned and governed.
    \item[G4] \emph{Portable verification}: any auditor can check the artifact without re-running the RM, and a contract can record a compact partial attestation for settlement together with the auditor decision.
    \item[G5] \emph{Bounded disclosure}: inputs, activations, and weights stay off the public path, and what auditors see is stated explicitly.
\end{itemize}
Sections~\ref{sec:artifact}--\ref{sec:pipeline} realize G1--G3 through artifact fields and binding checks, G4 through split verification, and G5 through the public-statement design analyzed in Section~\ref{sec:security}.

\subsection{Threat Model}
\label{sec:threat}
We consider a provider or intermediary that may return an incorrect reward, substitute a cheaper or outdated model checkpoint, alter the numerical configuration, replay an artifact under the wrong verifier, graft an accepted artifact onto another service request, or attach a valid-looking proof to a reward not derived from the proved execution. The adversary may also tamper with an exported package before it reaches an auditor or contract. The attack surface is the binding between score, request, model, verifier, configuration, and export, rather than the semantic quality of the RM itself. We assume standard cryptographic assumptions for the underlying proof and commitment schemes. We do not claim to prove that the RM itself is well trained, unbiased, or aligned. The \vrmaas{} framework provides \emph{service-output integrity}: the reported reward follows from the prescribed private computation under the declared artifact binding.

\section{Proof-Carrying Reward Artifacts}
\label{sec:artifact}

The central object in \vrmaas{} is a request-bound reward artifact that connects RM serving, independent audit, and optional public attestation across trust boundaries. This section defines the artifact fields, export semantics, proof-task interface, and verifier views used to check it.

\subsection{Artifact Fields and Export}
Let $M$ denote the prescribed reward-model service specification, including the architecture, numerical configuration, lookup tables, transcript rules, and verifier implementation, and let $W$ denote the private model weights. Each service invocation is bound to a public request identifier
\begin{equation}
\begin{aligned}
\mathsf{requestID}
&=
\mathsf{Hash}(
\mathsf{serviceID},
\mathsf{requesterID},
\mathsf{promptDigest},\\
&\hspace{2.2em}
\mathsf{responseDigest},
\mathsf{nonce},
\mathsf{policyID}),
\end{aligned}
\label{eq:request-id}
\end{equation}
\brev{Hash roles use disjoint domain tags and canonical field-element encodings: $\mathsf{Hash}$ derives request/object identifiers, $\mathsf{Hash}_{\mathrm{in}}$ derives salted prompt/response digests in off-chain request metadata, and $\mathsf{Digest}_{254}$ hashes canonical exports for off-chain audit. The service interface may receive either a raw prompt-response pair or a canonical RM-input package $I_{\mathrm{RM}}=(H^{(0)},M_{\mathrm{attn}},p_{\mathrm{last}})$ produced by the requester or a front-end adapter; the proof trace starts from these field-encoded hidden states. Request metadata identifies the invocation, while tokenization, embedding lookup, and the derivation of padding/segment metadata and the last-token position from raw text remain outside the implemented proof.} A conventional API returns only $r$ or a ranking decision. A \vrmaas{} service returns a proof-carrying artifact:
\begin{equation}
\begin{aligned}
\mathcal{A}_{\mathrm{off}}
=
\bigl(
&\mathsf{requestID},
r_{\mathrm{pub}},
\Pi_{\mathrm{comp}},\\
&\mathcal{C}_{W},
\mathcal{B}_{W},
\mathsf{verifierID},\mathsf{configID}
\bigr),
\end{aligned}
\label{eq:artifact}
\end{equation}
where $\mathsf{requestID}$ identifies the service invocation, $r_{\mathrm{pub}}$ is the public reward, $\Pi_{\mathrm{comp}}$ is a composite proof for the proof-relevant RM computation, $\mathcal{C}_{W}$ is a public commitment to the intended model weights, $\mathcal{B}_{W}$ is weight-binding material consumed by the exported verifier, and the two identifiers bind the artifact to a specific verifier implementation and numerical configuration.

For deployment, the off-chain artifact is serialized into a verifier-facing export:
\begin{equation}
\mathcal{E}
=
\mathsf{Export}(\mathcal{A}_{\mathrm{off}}),
\label{eq:export}
\end{equation}
which records the request identifier, \yrev{the digest-bearing request metadata}, public reward, deterministic proof-task order, compact proof encodings, the model-weight commitment, and the binding metadata needed by auditors or public verifiers. \rev{Table~\ref{tab:artifact-semantics} summarizes the service meaning of each field.}

The hidden witness contains the \brev{private canonical RM input package,} model weights, layer states, and auxiliary proof data:
\begin{equation}
\mathcal{W}
=
\left(
\brev{I_{\mathrm{RM}}},\,
W,\,
\{H^{(\ell)}\}_{\ell=0}^{L},\,
\omega_{\mathrm{aux}}
\right),
\end{equation}
where \brev{$I_{\mathrm{RM}}$ contains the initial field-encoded states, masks, and last-valid-token index supplied at the RM-service input boundary,} $\{H^{(\ell)}\}_{\ell=0}^{L}$ are the intermediate layer states, and $\omega_{\mathrm{aux}}$ collects auxiliary openings, randomness, and lookup witnesses used by the proof system. The verifier does not receive this witness. It receives a compact service artifact or exported package and checks whether the public reward is consistent with a valid hidden execution. All proof-side tensors are first mapped to finite-field values through the declared quantization and fixed-point rules in $M$. Thus, the object being verified is not an informal floating-point run, but a configured field-level execution of the RM service. The configuration identifier fixes the scale factors, lookup-table versions, range bounds, transcript format, and reward-export rule used to interpret the artifact.

\begin{table*}[t]
	\centering
	\scriptsize
	\setlength{\tabcolsep}{3pt}
	\caption{Service-contract semantics of \vrmaas{} artifact and statement fields. Complete binding checks are performed off chain; the public wrapper checks the supplied tuple and selected consistency relations in Section~\ref{sec:prototype-attestation}.}
	\label{tab:artifact-semantics}
	\begin{tabular}{p{0.12\textwidth}p{0.20\textwidth}p{0.27\textwidth}p{0.33\textwidth}}
		\toprule
		\textbf{Field} & \textbf{Verifier view} & \textbf{Binding target} & \textbf{Service-contract meaning} \\
		\midrule
		$\mathsf{requestID}$ &
		Public; export verifier and wrapper relation &
		\brev{Service ID, requester ID, digest-bearing request metadata, nonce, policy ID} &
		Names the accepted service job and prevents cross-request replay. \\
		$r_{\mathrm{pub}}$ &
		Public; composite proof and wrapper relation &
		RM execution under $M$ and $W$, last-token hidden state, reward-head output &
		Score or ranking value consumed by optimization, audit, moderation, or settlement services. \\
		$\Pi_{\mathrm{comp}}$ &
		Off-chain; \vrmaas{} verifier &
		Deterministic RM proof trace and ordered task descriptors &
		Provides computation integrity without exposing private inputs or activations. \\
		$\mathcal{C}_{W}$ &
		Public; opening verifier and registration check &
		Private weights $W$ and registered model checkpoint &
		Establishes model identity and prevents checkpoint substitution. \\
		$\mathcal{B}_{W}$ &
		\brev{Off-chain, auditor-visible; export verifier} &
		Proof-relevant weight openings under $\mathcal{C}_{W}$ &
		Links hidden proof operands to authenticated weights \brev{without publishing the full checkpoint}. \\
		$\mathsf{verifierID}$ &
		Registration and off-chain export verifier &
		Verifier implementation and transcript rules &
		Prevents verifier substitution and fixes the code path used to interpret the artifact. \\
		$\mathsf{configID}$ &
		Public; arithmetic, lookup, export, and wrapper checks &
		Quantization scales, lookup tables, range bounds, and reward-export rule &
		Prevents configuration swaps and makes quantized service semantics auditable. \\
		$\mathsf{opOrderHash}$ &
		Public; wrapper equality and off-chain order check &
		Declared deterministic proof-task order &
		Identifies the supplied task order in the public tuple; the auditor checks its correspondence to the trace. \\
		$\mathsf{H}_{E}$ &
		Off-chain; canonical export digest check &
		Canonical export $\mathcal{E}$ &
		Identifies the retained package for audit; omitted from the measured public tuple. \\
		\bottomrule
	\end{tabular}
\end{table*}

\begin{table}[t]
\centering
\scriptsize
\caption{\rev{Core notation. Table~\ref{tab:artifact-semantics} lists the service identifiers and artifact fields.}}
\label{tab:notation}
{
\begin{tabular}{@{}lp{0.65\columnwidth}@{}}
\toprule
Symbol & Meaning \\
\midrule
$M,W,I_{\mathrm{RM}}$ & Service specification, weights, input package \\
$n,d,L,d_k$ & Sequence length, hidden width, layer count, attention-key width \\
$X_\ell,H^{(\ell)}$ & Layer states in computation and witness notation \\
$\tau_i,m,\mathcal{T}$ & Task descriptor, task count, task families \\
$\nu,\ell_{\max}$ & Tensor variable count and its maximum \\
$\delta,B$ & Sumcheck degree bound and largest aggregation batch \\
$\mathbb{F},D_{\mathrm{KZG}}$ & Scalar field and supported KZG degree \\
$\mathcal{C}_W,\mathcal{B}_W$ & Weight commitment and opening material \\
$\Pi_{\mathrm{comp}},\mathcal{E}$ & Composite proof and canonical export \\
$V_{\vrmaas},\pi_G$ & Off-chain verifier and public wrapper proof \\
\bottomrule
\end{tabular}
}
\end{table}

\subsection{Proof Tasks}
The RM forward pass is decomposed into typed proof tasks rather than one monolithic Transformer proof. Each proof task is represented by a task descriptor $\tau_i$, which records both the operator being checked and the metadata needed to bind that operator to the service specification:
\begin{equation}
\tau_i=
\left(
\mathsf{type}_i,\ell_i,\mathsf{role}_i,\mathsf{shape}_i,
\mathsf{in}_i,\mathsf{out}_i,\mathsf{meta}_i
\right).
\label{eq:task-descriptor}
\end{equation}
Here $\mathsf{type}_i$ specifies the proof-task family, such as \textsf{matmul}, \textsf{add}, \textsf{mapping}, or \textsf{range}. The index $\ell_i$ locates the task in a Transformer layer or output stage, and $\mathsf{role}_i$ identifies its semantic role, for example query projection, attention-score computation, residual update, final token selection, or reward-head projection. The field $\mathsf{shape}_i$ fixes the tensor dimensions; $\mathsf{in}_i$ and $\mathsf{out}_i$ reference the committed or witness-side input and output values; and $\mathsf{meta}_i$ carries configuration-dependent metadata such as lookup-table identifiers, scale factors, range bounds, or weight-object identifiers. The service records the proof-relevant computation as a deterministic sequence of task descriptors:
\begin{equation}
\mathsf{Trace}(\brev{I_{\mathrm{RM}}},W,M)
=
(\tau_1,\tau_2,\ldots,\tau_m).
\label{eq:trace}
\end{equation}
Each proof task induces a relation
\begin{equation}
\mathcal{R}_{\tau_i}
\left(
\mathsf{pub}_{i};\mathsf{wit}_{i}
\right)=1,
\label{eq:task-relation}
\end{equation}
where $\mathsf{pub}_{i}$ contains the task descriptor, public commitments, and configuration-bound metadata, while $\mathsf{wit}_{i}$ contains the hidden tensor values and auxiliary proof openings.

Each task descriptor $\tau_i$ in the trace defined in \eqref{eq:trace} has a type field $\mathsf{type}_i$ satisfying $\mathsf{type}_i \in \mathcal{T}$, where $\mathcal{T}$ is the set of admissible proof-task families. Arithmetic tasks cover matrix multiplication, element-wise multiplication, addition, and equality checks used to bind registered values across the trace:
\begin{equation}
\mathcal{T}_{\mathrm{arith}}
=
\{
\textsf{matmul},
\textsf{hadamard},
\textsf{add},
\textsf{equality}
\}.
\label{eq:arith-tasks}
\end{equation}
For example, a matrix-multiplication task proves $C=AB$ over finite-field encodings by checking the multilinear-extension identity~\rev{\cite{cormode2012practical}}
\begin{equation}
\widetilde{C}(u,v)
=
\sum_{k\in\{0,1\}^{d_1}}
\widetilde{A}(u,k)\widetilde{B}(k,v),
\label{eq:matmul-relation}
\end{equation}
for verifier-chosen points through a sumcheck-style protocol. Addition, equality, and Hadamard tasks are treated as lower-degree tensor identities over the same task interface.

Lookup-backed tasks cover table-defined nonlinear maps and boundedness checks introduced by quantization and fixed-point rescaling:
\begin{equation}
\mathcal{T}_{\mathrm{lookup}}
=
\{
\textsf{range},
\textsf{mapping}
\}.
\label{eq:lookup-tasks}
\end{equation}
Mapping lookups prove that an input-output pair belongs to a configured public table, for example
\begin{equation}
(a,b)\in T_{\mathsf{map}}^{\mathsf{configID}},
\qquad b=f_{\mathsf{table}}(a),
\label{eq:mapping-lookup}
\end{equation}
where $T_{\mathsf{map}}^{\mathsf{configID}}$ encodes an elementwise scalar map such as the $\exp$ used in softmax, a reciprocal, an inverse-square-root for normalization, or a SwiGLU-style activation. Row-wise softmax and cross-dimension normalization are composed from these elementwise mapping and range tasks together with the \textsf{add}, \textsf{matmul}, and \textsf{hadamard} tasks that compute the row sums, means, and normalization-consistency checks. \brev{Softmax is therefore represented by exp lookup, row-sum, reciprocal, product-consistency, mask-consistency, and bounded-error checks, while RMS/LayerNorm-style normalization uses sum-of-squares or mean/variance tasks, inverse-square-root lookup, scale multiplication, and range/no-overflow checks.} Range lookups prove boundedness for quantized values, segment digits, intermediate integers, and fixed-point remainders:
\begin{equation}
z\in T_{\mathsf{range}}^{\mathsf{configID}}.
\label{eq:range-lookup}
\end{equation}
Range checks are therefore auxiliary numerical-semantics checks; they do not replace mapping lookups for nonlinear functions. Together, the admissible task-family set is
\begin{equation}
\mathcal{T}
=
\mathcal{T}_{\mathrm{arith}}
\cup
\mathcal{T}_{\mathrm{lookup}}.
\label{eq:task-family}
\end{equation}
This task-level interface is important for service-oriented reasoning: it makes explicit what the service promises to verify, what the artifact contains, and which parts of the private execution remain hidden.
The \vrmaas{} prototype realizes these tasks with standard cryptographic components: sumcheck-style arithmetic checks over multilinear tensor relations, lookup arguments for table-defined operations and range claims, polynomial commitments~\rev{\cite{kate2010constant}} with Gemini-style opening reduction~\rev{\cite{bootle2022gemini}} for tensor and weight openings, and a Groth16 wrapper for compact public attestation. Appendix~\ref{appx-app:proof-components} provides the technical background for these components.

\subsection{Verifier Views}
The off-chain verifier checks the registered primitive tasks induced by a claimed inference run. For arithmetic tasks, it invokes the corresponding sumcheck-style sub-verifiers. For table-defined nonlinear maps and auxiliary boundedness conditions, it invokes lookup verifiers. The public attestation path is smaller: it checks a compact statement rather than replaying every task. Section~\ref{sec:attestation} defines the implemented public tuple and partial wrapper relation, while Appendix~\ref{appx-app:extended-attestation} specifies an export-binding extension. This separation keeps local computation integrity, model-weight binding, and public consumption as distinct service interfaces.

\subsection{Service Lifecycle}
\label{sec:lifecycle}
\rev{\noindent\textbf{Registration.} Before serving, the model publisher registers $\mathcal{C}_W$, the verifier publisher registers $\mathsf{verifierID}$ with its verification and wrapper keys, and the deployment registers each $\mathsf{configID}$ with its admissibility decision (Section~\ref{sec:discussion-threats}). Consumers use these registrations to select the expected model, configuration, and verification material. The public path represents checkpoint registration through $\mathsf{Registered}(\cdot)$ under the assumptions in Section~\ref{sec:security}.}

\rev{\noindent\textbf{Invocation and delivery.} A requester submits its input with a fresh salted $\mathsf{nonce}$ and a $\mathsf{policyID}$, and the provider derives $\mathsf{requestID}$ by \eqref{eq:request-id}. The deployment can return $r_{\mathrm{pub}}$ and $\mathsf{requestID}$ immediately for provisional, non-critical use. Consequential actions, such as payment, an audit pass, or a governance decision, depend on an accepted artifact for that same $\mathsf{requestID}$. This interface accommodates the tens-of-seconds-to-minutes producer times in Section~\ref{sec:rq-cost}.}

\rev{\noindent\textbf{Audit and settlement.} Any party holding $\mathcal{E}$ can check its canonical digest and run $V_{\vrmaas}$ during an audit or dispute. For multi-party settlement, the prototype publishes $\mathrm{Stmt}^{\mathrm{proto}}_{\mathrm{pub}}$ and $\pi_G$ for the contract to record the partial statement. Complete acceptance combines this record with the auditor result; a complete contract proof would require binding $\mathsf{H}_E$ and verifying the complete relation.}

\rev{\noindent\textbf{Versioning.} Rotating a checkpoint, quantization configuration, or verifier produces new identifiers. Earlier artifacts remain verifiable under their original identifiers, so consumers can pin, deprecate, or reject versions without changing the meaning of historical rewards. These identifiers give service-version management an auditable basis beyond provider bookkeeping.}

\section{\vrmaas{} Service Pipeline}
\label{sec:pipeline}

Building on the overview in Fig.~\ref{fig:service-pipeline}, this section details the RM computation, composite proof construction, and weight-binding/export steps inside the serving pipeline. The pipeline has two coupled planes. The \emph{serving plane} computes the reward and records the proof-relevant execution trace. The \emph{verification plane} turns that trace into a composite proof, binds it to committed weights and declared verifier configurations, and exports a package that can be consumed by auditors or public attestation logic. Algorithm~\ref{alg:vrmaas-workflow} summarizes this end-to-end workflow.

\subsection{Reward-Model Computation}
The \vrmaas{} framework targets Transformer-based reward models used in alignment services. Fig.~\ref{fig:rm-proof-target} shows how the RM forward path is exposed as proof-task targets. The backbone computes contextual representations through stacked attention and feed-forward layers~\cite{vaswani2017attention,touvron2023llama}. The reward head maps the final representation of the last valid token to a scalar score:

\begin{figure*}[t]
	\centering
	\includegraphics[width=0.94\textwidth]{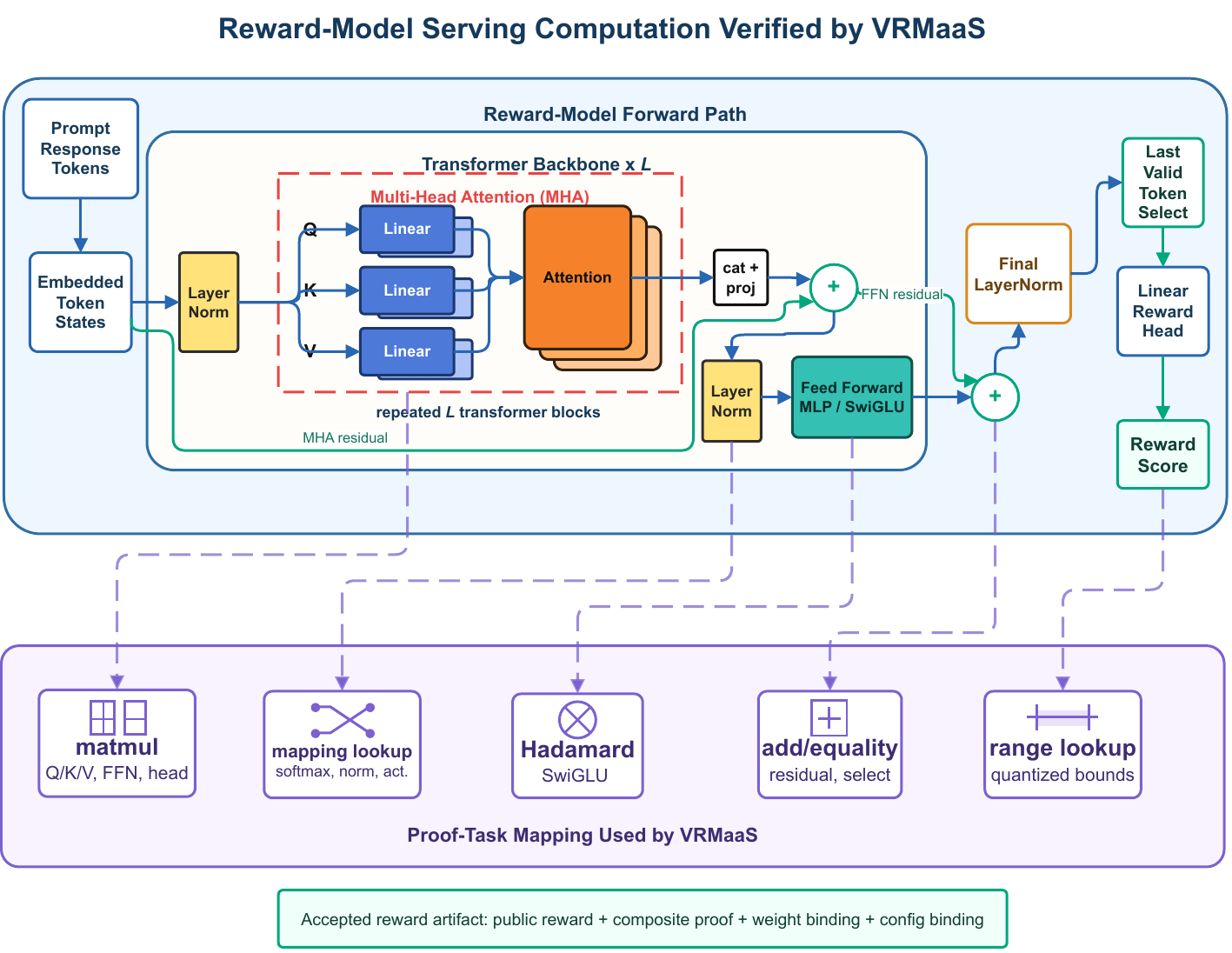}
	\caption{Reward-model computation verified by \vrmaas. The figure aligns the last-token Transformer RM forward path with the arithmetic and lookup-backed proof-task classes certified in the proof-carrying artifact.}
	\label{fig:rm-proof-target}
\end{figure*}

\brev{The canonical RM-input package supplies the initial field-encoded states, masks, and last-valid-token position used by the service invocation.} Let $X_{\ell-1}\in\mathbb{F}^{n\times d}$ denote the field-encoded hidden states entering Transformer layer $\ell$ after the quantization rules declared by $\mathsf{configID}$ have been applied. For clarity, we express the computation in standard tensor notation; the verifier interprets each operator as the corresponding fixed-point field relation and lookup table selected by the configuration. The attention sublayer is represented as
\begin{align}
\bar{X}_{\ell-1}
&= \mathsf{LN}_{\ell}^{\mathrm{att}}(X_{\ell-1}), \notag\\
Q_{\ell} &= \bar{X}_{\ell-1}W_{Q,\ell},\quad
K_{\ell} = \bar{X}_{\ell-1}W_{K,\ell},\quad
V_{\ell} = \bar{X}_{\ell-1}W_{V,\ell}, \notag\\
S_{\ell} &= Q_{\ell}K_{\ell}^{T}/\sqrt{d_k},\quad
A_{\ell} = \mathsf{Softmax}_{\mathsf{configID}}(S_{\ell}), \notag\\
O_{\ell} &= A_{\ell}V_{\ell}W_{P,\ell},\quad
R_{\ell}=X_{\ell-1}+O_{\ell}.
\label{eq:rm-attention-layer}
\end{align}
The feed-forward sublayer then applies pre-normalization, SwiGLU-style gating, and the second residual update:
\begin{align}
\bar{R}_{\ell}
&= \mathsf{LN}_{\ell}^{\mathrm{ffn}}(R_{\ell}), \notag\\
G_{\ell}
&= \psi_{\mathsf{configID}}
(\bar{R}_{\ell}W_{\mathrm{gate},\ell}),\quad
U_{\ell}=\bar{R}_{\ell}W_{\mathrm{up},\ell}, \notag\\
F_{\ell}
&=(G_{\ell}\odot U_{\ell})W_{\mathrm{down},\ell},\quad
X_{\ell}=R_{\ell}+F_{\ell}.
\label{eq:rm-ffn-layer}
\end{align}
Equations~\eqref{eq:rm-attention-layer} and~\eqref{eq:rm-ffn-layer} are not introduced as a new neural architecture. They define the service computation that the artifact certifies: every projection, attention product, nonlinear map, element-wise product, and residual update must appear in the proof trace under the declared model and numerical configuration.

\noindent
\begin{equation}
\begin{aligned}
h_{\mathrm{last}} &= X'_{L}[\texttt{last\_token\_pos}],\\
\widehat{r} &= h_{\mathrm{last}}^{T}W_{\mathrm{head}},\qquad
r_{\mathrm{pub}}=\mathsf{ExportReward}_{\mathsf{configID}}(\widehat{r}).
\end{aligned}
\label{eq:last-token-reward}
\end{equation}
The \vrmaas{} framework instantiates this last-valid-token scalar reward head in the evaluated service contract. Alternative pooling-based reward heads are outside the scope of that contract.

During serving, the provider does not expose the full tensor execution. Instead, it records a proof trace containing the ordered proof tasks induced by the RM forward pass. Each trace entry carries the operator type, position in the service configuration, tensor-shape metadata, and the commitment handles needed later by the verifier. This trace is deterministic for a fixed input, model specification, and configuration, which is essential for preventing ambiguity between the computation that produced the score and the computation that is later verified.

The trace follows the actual layer order of the RM rather than a generic neural-network template. For each Transformer layer $\ell$, the prover registers tasks for pre-attention normalization, query/key/value projections, attention-score computation, softmax-style normalization, value aggregation, output projection, residual addition, pre-FFN normalization, feed-forward projections, SwiGLU-style activation, Hadamard product, and the second residual addition. Linear projections and value aggregation become \textsf{matmul} tasks; residual updates become \textsf{add} tasks; element-wise products in the feed-forward block become \textsf{hadamard} tasks; and table-defined normalization, softmax, and activation maps become \textsf{mapping} tasks supported by \textsf{range} tasks for quantized bounds and remainders. After all $L$ blocks, the prover registers final normalization, last-valid-token selection, and the linear reward-head projection. Thus, the final scalar reward is not appended as an unproved API value: it is the output of the last registered proof task chain.

\begin{algorithm}[!t]
	\caption{\vrmaas{} Serving and Verification Workflow}
	\label{alg:vrmaas-workflow}
	\begin{algorithmic}[1]
		\Require \brev{Canonical RM input package $I_{\mathrm{RM}}$,} request identifier $\mathsf{requestID}$, service specification $M$, private weights $W$, registered commitment $\mathcal{C}_{W}$, identifiers $\mathsf{verifierID}$ and $\mathsf{configID}$
		\Ensure Proof-carrying artifact $\mathcal{A}_{\mathrm{off}}$ and optional public tuple $\mathrm{Stmt}^{\mathrm{proto}}_{\mathrm{pub}}$ and proof $\pi_G$
		\State Provider fixes $M$ under the declared verifier and numerical configuration
		\State $(r_{\mathrm{pub}},\mathsf{Trace}) \gets \mathsf{ServeRM}(\brev{I_{\mathrm{RM}}},W,M)$
		\State Compile $\mathsf{Trace}$ into typed tasks $\tau_i \in \mathcal{T}$
		\State Generate arithmetic and lookup-backed subproofs for the recorded tasks
		\State Aggregate ordered subproofs into $\Pi_{\mathrm{comp}}$
		\State Prepare $\mathcal{B}_{W}$ by opening proof-relevant weight values against $\mathcal{C}_{W}$
		\State Form $\mathcal{A}_{\mathrm{off}}$ as in \eqref{eq:artifact}
		\State Export $\mathcal{E}\gets\mathsf{Export}(\mathcal{A}_{\mathrm{off}})$
		\State Compute $\mathsf{H}_{E}\gets\mathsf{Digest}_{254}(\mathcal{E})$ for audit
		\State Auditor checks $V_{\vrmaas}(\mathcal{E})=1$ under the declared request and configuration
		\If{public attestation is requested}
		\State Form $\mathrm{Stmt}^{\mathrm{proto}}_{\mathrm{pub}}$ as in \eqref{eq:public-statement}
		\State Generate wrapper proof $\pi_G$ for $\mathcal{R}_{\mathrm{wrap}}$
		\State Contract returns $\mathsf{Accept}_{\mathrm{chain}}$
		\EndIf
	\end{algorithmic}
\end{algorithm}

\subsection{Composite Proof Construction}
During inference, the \vrmaas{} framework uses the deterministic proof trace in \eqref{eq:trace}, where each descriptor $\tau_i$ belongs to a configured task family $\mathcal{T}$. Each task is checked by the corresponding arithmetic or lookup-based verifier. The resulting composite proof packages the ordered trace commitment and the subproofs needed by the exported verifier:
\begin{equation}
\Pi_{\mathrm{comp}}
=
\left(
\mathsf{op\_order},
\Pi_{\mathrm{mat}},
\Pi_{\mathrm{add}},
\Pi_{\mathrm{had}},
\Pi_{\mathrm{range}},
\Pi_{\mathrm{map}}
\right).
\label{eq:composite-proof}
\end{equation}
The deterministic order binds the proof to the declared RM execution trace and to the transcript used for Fiat--Shamir challenges. In the prototype, these challenges are derived from a BN254 Poseidon2 transcript. The prover and verifier absorb the ordered task descriptors, commitments, and proof messages in the same deterministic order, so arithmetic, lookup, and commitment-opening checks are bound to the same proof trace. This transcript role is distinct from the export digest described in Section~\ref{sec:attestation}: the transcript derives proof challenges, whereas the digest identifies a canonical package for off-chain audit. Binding that digest inside a public wrapper is the additional relation specified in Appendix~\ref{appx-app:extended-attestation}. The exported verifier interprets each subproof under the declared $\mathsf{op\_order}$ and $\mathsf{configID}$; a proof generated for one operator order, lookup-table version, or quantization configuration cannot be silently reinterpreted as another. This is the main reason the \vrmaas{} framework treats proof-task order and configuration identifiers as part of the service artifact rather than as local implementation details.

Matrix multiplications dominate Transformer reward serving, so the \vrmaas{} framework aggregates same-shaped matmul tasks instead of proving each task in isolation. Suppose $N$ registered tasks claim $C^{(n)}=A^{(n)}B^{(n)}$ for $n\in[N]$. Let $\widetilde{A}^{(n)}$, $\widetilde{B}^{(n)}$, and $\widetilde{C}^{(n)}$ denote multilinear extensions of the encoded tensors, and let $\lambda_n(\rho)$ be transcript-derived random aggregation weights. The verifier checks one aggregated relation:
\begin{equation}
\sum_{n=1}^{N}
\lambda_n(\rho)
\left(
\widetilde{C}^{(n)}(u,v)
-
\sum_{k}
\widetilde{A}^{(n)}(u,k)
\widetilde{B}^{(n)}(k,v)
\right)
=0.
\label{eq:matmul-aggregation}
\end{equation}
If every registered matmul task is correct, \eqref{eq:matmul-aggregation} holds. If any task is false, the random linear combination is nonzero except with Schwartz--Zippel probability over the verifier challenges~\cite{schwartz1980fast,zippel1979probabilistic}. The same exported proof context records which task descriptors are included in the aggregation batch, so a prover cannot silently move a projection, reward-head multiplication, or attention multiplication into a different batch.
In the prototype, the resulting multilinear-opening claims are discharged through Gemini folding into polynomial-commitment openings.

Lookup-backed tasks are aggregated at the level of table membership and boundedness claims using the same lookup interface. A \textsf{mapping} task contributes committed input-output pairs to the lookup argument for the table identified by $\mathsf{configID}$, while \textsf{range} tasks contribute bounded values for quantization segments and fixed-point remainders. The verifier therefore checks not only that a nonlinear map was satisfied for private values, but also that the values were interpreted under the declared table version and numerical range. This is the technical link between proof soundness and service configuration: changing the table, scale, or range changes the verified relation.

\subsection{Weight Binding and Export}
A valid computation proof must not be detachable from the model checkpoint it purports to use. Before export, the provider prepares weight-binding material for the proof-relevant weights:
\begin{equation}
\mathcal{B}_{W}
=
\{(\mathsf{id}_{j},z_j,v_j,\pi_j)\}_{j=1}^{q},
\label{eq:weight-binding-material}
\end{equation}
where $\mathsf{id}_{j}$ identifies a proof-relevant weight entry or committed polynomial, $z_j$ is the opened point, $v_j$ is the claimed value, and $\pi_j$ is the corresponding opening proof. A weight identifier is not merely an array offset. It is derived from the service specification, layer index, tensor role, quantization view, and committed object identifier, e.g.,
\begin{equation}
\mathsf{id}_j
=
\mathsf{Hash}
(\mathsf{modelID},\ell,\mathsf{role},\mathsf{shape},\mathsf{view}).
\label{eq:weight-object-id}
\end{equation}
This identifier is referenced both by the proof task that consumes the value and by the commitment-opening verifier. The \vrmaas{} framework checks these openings against the registered model-weight commitment:
\begin{equation}
\forall j\in[q],\quad
\mathsf{VerifyOpen}_{W}
\left(
\mathcal{C}_{W},
\mathsf{id}_{j},
z_{j},
v_{j},
\pi_{j}
\right)=1.
\label{eq:weight-open-check}
\end{equation}
The exported verifier also checks that the value authenticated by \eqref{eq:weight-open-check} is the same value referenced by the corresponding proof task:
\begin{equation}
\forall (i,j)\in\mathcal{L},\quad
\mathsf{WeightRef}(\tau_i,\mathsf{id}_j)=v_j,
\label{eq:weight-task-link}
\end{equation}
where $\mathcal{L}$ is the link set between task descriptors and opened weight objects. Equations~\eqref{eq:weight-open-check} and~\eqref{eq:weight-task-link} are the two sides of model binding: the first authenticates a value against $\mathcal{C}_{W}$, while the second proves that the authenticated value is the one actually used by the RM proof task. Therefore, a prover cannot generate a valid computation proof with one set of weight values and later attach it to a different model commitment, except by breaking the commitment binding or the soundness of the relevant proof task. \brev{The opening value $v_j$ is part of the auditor-facing verification material: an off-chain auditor that checks weight binding observes per-request polynomial evaluation values, whereas the public statement carries only $\mathcal{C}_{W}$ and not the full checkpoint or opening transcript.}
The export step follows \eqref{eq:export}. The request identifier binds the package to the service invocation, while the verifier and configuration identifiers bind the package to a verifier implementation and to the quantization, lookup-table, transcript, and reward-export configuration used by that verifier. The result is a portable service artifact: an auditor can verify the request-bound reward without rerunning the private RM, and a public attestation layer can consume a compact statement derived from the same export.

\subsection{Efficient Artifact Generation}
\label{sec:producer-acceleration}
We optimize witness construction and weight binding, two major sources of producer work, while preserving the arithmetic relations and the weight-to-task links in Equations~\eqref{eq:weight-open-check}--\eqref{eq:weight-task-link}. The first optimization uses the bounded integer values of quantized tensors; the second combines opening claims within a request after the transcript has fixed the relevant objects. Lookup-workspace reuse and specialized evaluation kernels reduce allocation and polynomial-evaluation overhead around these two paths.

\noindent\textbf{Exact integer witnesses.}
For a dot product of length $K$ with integer operands bounded by $A_{\max}$ and $B_{\max}$, the producer checks that $K A_{\max}B_{\max}<2^{63}$ and that the signed result has a unique representation in the proof field. It then accumulates in signed 64-bit integers and converts the result to the field representation expected by the proof tasks. This replaces expensive field operations in witness construction without changing the corresponding field-arithmetic relation. The optimization applies only to numerical configurations whose range bounds satisfy these conditions. Reusing lookup buffers further avoids repeated workspace allocation during trace processing and proving.

\noindent\textbf{Request-level opening aggregation.}
The producer first absorbs the ordered weight-object descriptors and commitments into the transcript, then derives a shared Gemini challenge. Transcript-derived coefficients combine KZG opening claims that use the same evaluation point, so the prover generates one combined opening for that point rather than separate openings for every object. The verifier checks the matching linear combination of commitments and evaluations. The construction retains the per-object links between authenticated weight values and the computation tasks; it reduces opening work within the current request. Evaluation points, coefficients, and opening proofs depend on that request's transcript and are generated anew for each artifact.

\noindent\textbf{MSM and evaluation kernels.}
For the low-entropy scalars of quantized weights, the prover uses scalar whitening to improve the workload distribution of bucket-based multi-scalar multiplication (MSM). For a nonzero random field element $\alpha$, it computes the MSM with scalars $\alpha s_i$ and rescales the group result by $\alpha^{-1}$; linearity leaves the commitment $\sum_i s_i P_i$ unchanged. This is a performance transformation, not an additional hiding guarantee. A parallel Horner kernel handles evaluation-only calls to $f(z)$ without constructing an unused quotient polynomial. Calls that produce a KZG opening still generate the opening proof. Static SRS material, checkpoint commitments, and configuration metadata may be retained across requests; transcript-dependent opening generation remains part of the measured per-request cost.

\section{Public Wrapper Attestation}
\label{sec:attestation}
Optional public attestation gives downstream contracts a compact proof of selected consistency checks and supplied service fields. The complete exported verifier $V_{\vrmaas}(\mathcal{E})$ checks computation, model-weight openings, and service bindings off chain. The public path records a partial attestation that can be used together with that auditor decision in a settlement workflow. Keeping the two acceptance rules explicit lets consumers distinguish the evidence supplied by the artifact from the checks performed by the contract.

\subsection{Export Digest and Public Statement}
The producer canonically serializes the artifact and computes its BN254-native Poseidon2 digest:
\begin{equation}
\begin{aligned}
\mathsf{Digest}_{254}(\mathcal{E})
&=\mathsf{Poseidon2}_{\mathrm{BN254}}(\mathsf{Encode}(\mathcal{E})),\\
\mathsf{H}_{E}&=\mathsf{Digest}_{254}(\mathcal{E}).
\end{aligned}
\label{eq:export-digest}
\end{equation}
Auditors use $\mathsf{H}_{E}$ to identify and check the canonical package. This export digest has a different role from the Fiat--Shamir transcript, which derives challenges from ordered task descriptors, commitments, and proof messages. The measured public wrapper exposes the following tuple:
\begin{equation}
\mathrm{Stmt}^{\mathrm{proto}}_{\mathrm{pub}}=
\bigl(\mathsf{requestID},r_{\mathrm{pub}},\mathcal{C}_{W},
\mathsf{configID},\mathsf{opOrderHash}\bigr).
\label{eq:public-statement}
\end{equation}
Here $\mathsf{opOrderHash}$ represents the declared proof-task order. The tuple supplies the request, reward, checkpoint commitment, configuration, and task-order hash to the wrapper; its field equalities bind a wrapper proof to those supplied values. The export digest $\mathsf{H}_{E}$ and $\mathsf{verifierID}$ remain off-chain fields, with the verifier implementation and wrapper keys selected through registration. Thus, matching the tuple to the complete exported computation remains part of the auditor's acceptance rule.

\subsection{Implemented Partial Public Attestation}
\label{sec:prototype-attestation}
The implemented Groth16 circuit~\cite{groth2016} checks selected Gemini fold values, partial sumcheck data, a reward claim, and private copies of the public fields. Its relation is
\begin{equation}
\begin{aligned}
&\mathcal{R}_{\mathrm{wrap}}:\ \exists\,w_{\mathrm{wrap}}\ \text{such that}\\
&\quad\mathsf{FoldCheck}(w_{\mathrm{wrap}})=1,\quad
\mathsf{SumcheckPartial}(w_{\mathrm{wrap}})=1,\\
&\quad\mathsf{RewardClaim}(w_{\mathrm{wrap}})=r_{\mathrm{pub}},\\
&\quad\mathsf{PublicFields}(w_{\mathrm{wrap}})=\\
&\qquad(\mathsf{requestID},\mathcal{C}_{W},\mathsf{configID},\mathsf{opOrderHash}).
\end{aligned}
\label{eq:wrapper-relation}
\end{equation}
The prover uses a circuit $C_{\mathrm{wrap}}$ implementing this relation to generate
\begin{equation}
\pi_G\leftarrow\mathsf{Groth16.Prove}
\bigl(C_{\mathrm{wrap}},\mathrm{Stmt}^{\mathrm{proto}}_{\mathrm{pub}};w_{\mathrm{wrap}}\bigr).
\label{eq:groth16-wrapper-proof}
\end{equation}
The circuit does not derive all public fields from the complete export, recursively verify the complete composite proof, or bind separate KZG calldata to the same private opening values. These distinctions specify the accepted relation and the scope of the gas and proof-size measurements. Appendix~\ref{appx-app:extended-attestation} retains an export-binding extension and states the additional constraints required to realize it.

\subsection{Contract Acceptance and Audit}
The contract receives $\mathrm{Stmt}^{\mathrm{proto}}_{\mathrm{pub}}$ and $\pi_G$. Under authenticated checkpoint registration, its acceptance rule is
\begin{equation}
\begin{aligned}
\mathsf{Accept}_{\mathrm{chain}}
=&\ \mathsf{Registered}(\mathcal{C}_{W})\\
&\wedge\mathsf{Verify}_{\mathrm{Groth16}}
(\mathsf{vk}_{G},\mathrm{Stmt}^{\mathrm{proto}}_{\mathrm{pub}},\pi_{G}).
\end{aligned}
\label{eq:chain-accept}
\end{equation}
After acceptance, the contract can return the reward associated with the supplied tuple or emit an attestation event for subsequent audit. The event records satisfaction of Equation~\eqref{eq:wrapper-relation}; it does not replace complete artifact verification. A consumer requiring the complete guarantee checks the auditor result for the same request, model, configuration, reward, and task order before acting on this record. This interface keeps the checkpoint, hidden activations, opening values, and complete inner transcript outside the public contract inputs.

\section{Security and Service Integrity}
\label{sec:security}

The \vrmaas{} framework provides service-output integrity against a malicious polynomial-time provider that may choose arbitrary private inputs, hidden tensors, proof tasks, exported packages, and wrapper witnesses in an attempt to make an auditor or contract accept an invalid reward. We assume that arithmetic subprotocols are sound over the configured field, lookup arguments are sound for the configured mapping and range tables, commitment openings are binding with respect to the registered model-weight commitment $\mathcal{C}_{W}$, the Poseidon2 transcript used for Fiat--Shamir challenges is modeled as a domain-separated random oracle, $\mathsf{Digest}_{254}$ is collision resistant for canonical encodings of exported packages, and the Groth16 wrapper is complete, sound, and zero knowledge for the partial wrapper relation \eqref{eq:wrapper-relation}. We also assume that $\mathsf{requestID}$ identifies the intended service invocation, that $\mathsf{verifierID}$ and $\mathsf{configID}$ identify the intended verifier implementation, quantization rules, lookup tables, and transcript format, and that the deployment accepts $\mathcal{C}_{W}$ only when it is registered or otherwise authenticated for the intended RM service. The off-chain and public acceptance statements below apply to their respective implemented relations; Appendix~\ref{appx-app:extended-attestation} separately describes the export-binding extension. The resulting risk-to-guarantee mapping is expanded in Appendix Table~\ref{appx-tab:threat-guarantee}.

\begin{lemma}[Completeness of accepted exports]
\label{lem:export-completeness}
If the provider honestly runs the prescribed RM, records the deterministic proof trace, generates valid arithmetic and lookup subproofs, prepares valid weight-binding material for $\mathcal{C}_{W}$, and exports $\mathcal{E}$ under the intended $\mathsf{requestID}$, $\mathsf{verifierID}$, and $\mathsf{configID}$, then \brev{$V_{\vrmaas}(\mathcal{E})=1$ for the off-chain verifier.} If public attestation is used and an honest wrapper witness satisfies \eqref{eq:wrapper-relation}, the Groth16 verifier accepts \eqref{eq:chain-accept}, except with negligible proof-system failure probability.
\end{lemma}

\begin{proof}[Proof sketch]
For an honest run, every task descriptor is generated from the prescribed RM trace and every arithmetic, lookup, and weight-opening claim is generated from the corresponding witness values. Completeness of the arithmetic subprotocols, lookup arguments, and polynomial-commitment openings therefore makes the off-chain verifier accept under the declared request and configuration. Groth16 completeness makes the contract accept an honest wrapper witness that satisfies \eqref{eq:wrapper-relation}.
\end{proof}

\begin{lemma}[RM-computation soundness]
\label{lem:rm-soundness}
If
\(
\brev{V_{\vrmaas}(\mathcal{E})=1,}
\)
then, except with negligible probability, the accepted package contains a valid computation proof for the claimed RM service execution. In particular, the arithmetic proof tasks, mapping lookups, and range lookups induced by the deterministic proof trace are satisfied; the public reward is derived from the proved last-token hidden state and reward head; and the proof-relevant weight values are bound to $\mathcal{C}_{W}$.
\end{lemma}

\begin{proof}[Proof sketch]
The verifier checks $\Pi_{\mathrm{comp}}$ against the ordered proof-task descriptors and checks $\mathcal{B}_{W}$ against the registered commitment. A false arithmetic, mapping, or range task would violate the soundness of the corresponding subprotocol. The wiring consistency that composes these per-task checks into a single end-to-end statement, together with the union bound over all $m$ task-level soundness errors, is formalized in Appendix~\ref{appx-app:composition}. A mismatch between opened proof-relevant weights and $\mathcal{C}_{W}$ would violate commitment binding. Since the reward head is the final registered task chain, an accepted $r_{\mathrm{pub}}$ is tied to the proved RM execution rather than appended as an unproved API value.
\end{proof}

Let $\delta$ bound the individual degree of each sumcheck polynomial, $\ell_{\max}$ the maximum number of variables in a committed tensor handle, and $B$ the largest aggregation batch. For $m$ tasks, the composition argument in Appendix~\ref{appx-app:composition} yields
\begin{equation}
\epsilon_{\mathrm{comp}}\le
\frac{c\,(m\,\ell_{\max}\,\delta+B)}{|\mathbb{F}|}
+\epsilon_{\mathrm{bind}}+\frac{Q}{2^{\kappa}},
\label{eq:composite-bound}
\end{equation}
where $c$ counts the arithmetic, lookup, wiring, and folding checks per task, $\epsilon_{\mathrm{bind}}$ is the commitment-binding advantage, and the domain-separated Fiat--Shamir transcript has $\kappa$-bit output and at most $Q$ oracle queries. The appendix retains the wiring lemma and full composition argument. For the prototype parameters ($m\leq 2^{15}$, $\ell_{\max}\leq24$, $\delta\leq3$, $B\leq2^{10}$), the algebraic field-checking term is below $2^{-220}$; overall security also depends on the computational BN254 binding assumption and the Fiat--Shamir term, so this is not a 220-bit system-security claim.

\begin{theorem}[Partial public attestation]
\label{thm:public-attestation-soundness}
If the contract accepts \eqref{eq:chain-accept}, then, except with negligible probability, there exists a wrapper witness satisfying the selected Gemini fold, partial sumcheck, reward-consistency, and public-field equality checks in \eqref{eq:wrapper-relation}. This theorem does not imply acceptance of the complete off-chain verifier $V_{\vrmaas}$ or of the separate KZG calldata for the same private openings.
\end{theorem}

\begin{proof}[Proof sketch]
Groth16 knowledge soundness yields $w_{\mathrm{wrap}}$ satisfying \eqref{eq:wrapper-relation}. The public-input equalities prevent a proof produced for one supplied tuple from being replayed under a different tuple without a new valid wrapper proof. The present relation does not, however, derive those fields from the complete export or recursively verify the off-chain KZG and composite-proof checks.
\end{proof}

\begin{lemma}[Public-attestation privacy boundary]
\label{lem:public-privacy}
Under the zero-knowledge property of the Groth16 wrapper, the public attestation path reveals no information about $w_{\mathrm{wrap}}$ beyond the truth of \eqref{eq:wrapper-relation} and the public fields in $\mathrm{Stmt}^{\mathrm{proto}}_{\mathrm{pub}}$.
\end{lemma}

\begin{proof}[Proof sketch]
The contract receives only $\mathrm{Stmt}^{\mathrm{proto}}_{\mathrm{pub}}$ and $\pi_G$. The statement contains the request identifier, reward, model commitment, configuration identifier, and task-order hash; the prompt, response text, hidden activations, model weights, opening values, and inner proof transcript are not public wrapper inputs. Zero knowledge hides the witness used to establish the selected relation, while the public fields retain their intended disclosure. Privacy against guessing the request from its identifier additionally assumes that the nonce is a high-entropy secret salt and the digest-bearing request metadata remains off chain. An off-chain auditor that verifies $\mathcal{B}_{W}$ observes the per-request weight-evaluation values, so hiding those values would require a hiding commitment or zero-knowledge opening variant.
\end{proof}

The composite-soundness statement establishes integrity of an export accepted by the complete off-chain verifier; the public theorem establishes the selected wrapper relation. Both concern the declared computation and service fields. Fairness, bias, robustness to reward hacking, and semantic reward quality remain properties to evaluate in the selected RM, as discussed in Section~\ref{sec:discussion-threats}.

\section{Experimental Evaluation}
\label{sec:evaluation}

The \vrmaas{} framework is evaluated as a proof-carrying service system rather than as a new reward-model training method. The evaluation asks whether an RM service can preserve baseline scoring behavior, export request-bound artifacts, support independent verification, and expose provider, auditor, and settlement-layer costs. We use three evidence types: direct prototype measurements (\textsc{D}), real-artifact verifier or contract replay (\textsc{R}), and trace-driven service simulation (\textsc{S}). This separation keeps artifact-producing measurements distinct from deployment-policy replays built over measured stage latencies. The subsections evaluate service fidelity (RQ1), proof-aware serving cost (RQ2), \brev{public wrapper attestation} (RQ3), service-bound rejection (RQ4), service operation (RQ5), design necessity and generic baselines (RQ6), and configuration limits (RQ7).
\subsection{Setup}
We evaluate a \vrmaas{} prototype on Skywork 1B, 3B, and 8B reward models~\cite{liu2025skywork} at sequence lengths 128--1024 using repeated question-answer comparisons. For each setting, we compare the proof-aware and baseline scoring paths and record fidelity, latency, proof-stage cost, binding cost, \brev{public-attestation cost}, payload size, proof size, and rejection behavior. The prototype uses CUDA components for RM inference and proof-side kernels, with scripts for model loading, input preparation, and deployment orchestration. Experiments run on a node with a 192-core AMD EPYC 9654 CPU, 503.35 GB RAM, and an NVIDIA A100 GPU with 80 GB memory.

The metrics are chosen to match the service contract defined earlier. We separate proof-fidelity metrics from benchmark-label behavior. Let $r^{\mathrm{base}}$ and $r^{\mathrm{proof}}$ denote baseline and proof-aware scalar rewards. For a chosen/rejected pair, define $\Delta^{p}=r^{p}_{c}-r^{p}_{r}$ for path $p\in\{\mathrm{base},\mathrm{proof}\}$. The score error, baseline-proof pairwise agreement, and benchmark pair accuracy are
\begin{equation}
\begin{aligned}
\mathrm{ScoreErr}
&=\left|r^{\mathrm{proof}}-r^{\mathrm{base}}\right|,\\
\mathrm{PairAgree}
&=\mathbf{1}\!\left[
\operatorname{sign}(\Delta^{\mathrm{proof}})
=
\operatorname{sign}(\Delta^{\mathrm{base}})
\right],\\
\mathrm{BenchAcc}^{p}
&=\mathbf{1}\!\left[\Delta^{p}>0\right].
\end{aligned}
\label{eq:fidelity-metrics}
\end{equation}
Here $\mathrm{PairAgree}$ measures baseline-proof service fidelity, while $\mathrm{BenchAcc}$ measures benchmark-label RM behavior and is not proof-system fidelity. Producer timing separates the integer forward pass, trace/preprocessing, computation proving (P2), producer-side verification (P3), weight binding (P4), and export serialization. These stage names also identify the historical records retained in the appendices. Before- and after-optimization measurements are reported separately because the implementations and execution settings differ; the optimized proof runs include attention-score and value-aggregation products. Deployment metrics cover gas, payload size, Groth16 proof size, calldata size, request-binding overhead, and modified-artifact rejection. Appendix Table~\ref{appx-tab:eval-matrix} records the full evaluation matrix behind these service-facing metrics.

\subsection{RQ1: Service Fidelity}
RQ1 asks whether proof-aware serving preserves the service output of the baseline RM path. Table~\ref{tab:service-fidelity} gives the main fidelity results; Appendix Table~\ref{appx-tab:benchmark-context} and Fig.~\ref{appx-fig:pairwise-agreement-trend} report benchmark preference accuracy and pairwise-agreement trends as RM-quality context.

\begin{table}[t]
\centering
\scriptsize
\caption{Service fidelity between the proof-aware path and the baseline RM path. Evidence: \textsc{D}. Pairwise agreement is measured against the baseline service decision; each setting uses $n=200$ deterministic chosen/rejected comparisons.}
\label{tab:service-fidelity}
\resizebox{\columnwidth}{!}{%
\begin{tabular}{llllllll}
\toprule
Model & Seq & Score R2 & MAE & RMSE & Pairwise Agreement & Flip Rate & Mean Quant Err \\
\midrule
1B & 128 & 0.9959 & 0.1537 & 0.1969 & 98\% \brev{(196/200)} & 2\% & 0.1537 \\
1B & 256 & 0.9860 & 0.3142 & 0.4007 & 97.50\% \brev{(195/200)} & 2.50\% & 0.3142 \\
1B & 512 & 0.9630 & 0.5682 & 0.7630 & 96.50\% \brev{(193/200)} & 3.50\% & 0.5682 \\
1B & 1024 & 0.9067 & 0.9094 & 1.2212 & 93.50\% \brev{(187/200)} & 6.50\% & 0.9094 \\
3B & 128 & 0.9979 & 0.0889 & 0.1220 & 99\% \brev{(198/200)} & 1\% & 0.0889 \\
3B & 256 & 0.9917 & 0.2268 & 0.3607 & 98\% \brev{(196/200)} & 2\% & 0.2268 \\
3B & 512 & 0.9619 & 0.7421 & 1.0763 & 97.50\% \brev{(195/200)} & 2.50\% & 0.7421 \\
3B & 1024 & 0.8849 & 1.7303 & 2.3360 & 95\% \brev{(190/200)} & 5\% & 1.7303 \\
8B & 128 & 0.9989 & 0.1062 & 0.1497 & 97.50\% \brev{(195/200)} & 2.50\% & 0.1062 \\
8B & 256 & 0.9963 & 0.2391 & 0.3821 & 97.50\% \brev{(195/200)} & 2.50\% & 0.2391 \\
8B & 512 & 0.9745 & 0.9388 & 1.4023 & 96\% \brev{(192/200)} & 4\% & 0.9388 \\
8B & 1024 & 0.9118 & 2.3256 & 3.2759 & 94\% \brev{(188/200)} & 6\% & 2.3256 \\
\bottomrule
\end{tabular}%
}
\end{table}

Across the 12 settings, pairwise agreement ranges from 93.50\% to 99\%, and score-path $R^2$ remains high for short and medium contexts. The lower $R^2$ values at 1024 tokens, especially for 3B and 8B, are consistent with greater long-context quantization drift under the registered fixed-point configuration. The two paths retain the same RM objective, so the comparison evaluates numerical fidelity rather than a change in the learning objective. This is why the verifier-facing $\mathsf{configID}$ is not cosmetic metadata: deployments can register, warn on, or reject numerical configurations based on observed drift thresholds. Appendix Table~\ref{appx-tab:case-level-fidelity} and the benchmark-label context in Appendix~\ref{appx-app:additional-fidelity} show why benchmark correctness and proof-path fidelity are reported separately: a proof-carrying artifact can preserve a reward service's decision even when the underlying RM disagrees with the benchmark label.

\subsection{RQ2: Proof-Aware Serving Cost}
\label{sec:rq-cost}
\begin{figure}[t]
  \centering
  \includegraphics[width=\columnwidth]{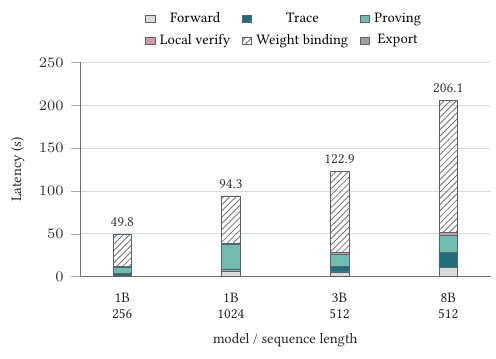}
  \caption{\rev{Optimized producer-side latency by stage, with end-to-end totals in seconds. Evidence: \textsc{D}. The four bars average 3, 1, 5, and 3 completed runs with per-request weight binding, respectively. Forward denotes integer RM execution within each artifact-producing run.}}
  \label{fig:latency-breakdown}
\end{figure}

\rev{Fig.~\ref{fig:latency-breakdown} reports the stage breakdown of artifact generation after the witness-construction and weight-binding optimizations. End-to-end generation takes 49.8 seconds for 1B/256, 94.3 seconds for 1B/1024, 122.9 seconds for 3B/512, and 206.1 seconds for 8B/512. Across these settings, the integer forward pass takes 1.28--11.10 seconds, trace/preprocessing 2.25--16.81 seconds, computation proving 6.74--28.23 seconds, producer-side verification 1.40--2.66 seconds, weight binding 37.87--154.63 seconds, and export 0.04--0.18 seconds. Weight binding remains the largest producer-side component. Separately measured weight binding for 8B/1024 averages 198.30 seconds; this is a stage time, not an end-to-end total. The attention-score $QK^\top$ and value-aggregation products are included in these artifact-producing proof runs.}

\rev{Component profiling reports weight-binding time reductions of 24--37\% for shared-challenge opening aggregation, 30--50\% for scalar whitening, and 17.5--26\% for evaluation-only Horner kernels. Integer matmul accelerates the forward stage by $2.8$--$4.1\times$ with identical reward outputs in the reported comparisons. These measurements use representative settings and profiling conditions; the percentages are component-level comparisons and are not combined into an end-to-end speedup. Public attestation is optional: an off-chain-only deployment omits the wrapper and pays no gas. The measured 49.8--206.1 second artifact-generation times support delayed audit, settlement, and governance.}

\rev{\emph{Before and after optimization.} Table~\ref{tab:binding-history} reports weight-binding times before and after optimization at matching model sizes and sequence lengths. Before optimization, the 8B/1024 run spent 3802.00 seconds on weight binding and 5541.75 seconds on artifact generation overall. After optimization, the 8B/1024 weight-binding mean is 198.30 seconds. The records show how measured weight-binding cost changed between the two implementations. The comparison is at the version level: the computation proof after optimization also includes attention products added after the pre-optimization measurements, and profiling and execution settings differ between versions. The stage times describe observed before/after costs without attributing the entire difference to a controlled optimization ablation.}

\begin{table}[t]
\centering
\scriptsize
\setlength{\tabcolsep}{4pt}
\caption{\rev{Weight-binding stage times before and after optimization at matching model sizes and sequence lengths. Counts give completed runs before/after optimization; the accompanying text describes the differences in proof coverage and execution settings.}}
\label{tab:binding-history}
{
\begin{tabular}{@{}lrrr@{}}
\toprule
Model/Seq & \shortstack{Before\\optimization (s)} & \shortstack{After\\optimization (s)} & Runs \\
\midrule
1B/128 & 1156.53 & 37.68 & 41/5 \\
3B/512 & 2319.00 & 94.32 & 1/5 \\
8B/1024 & 3802.00 & 198.30 & 1/5 \\
\bottomrule
\end{tabular}
}
\end{table}

Table~\ref{tab:optimized-stages} gives the numerical stage totals underlying the optimized latency plot. The original complete-run breakdown is retained in Appendix~\ref{appx-app:historical-latency}. That historical baseline and the no-proof forward measurements in Appendix Table~\ref{appx-tab:baseline-latency-revised} are not matched executions of the optimized implementation, so their difference is not used as a controlled end-to-end speedup estimate.

\begin{table*}[t]
\centering
\small
\caption{Complete artifact generation after optimization. All times are seconds; each row includes fresh per-request weight-binding openings. Runs give completed artifact-producing executions. Forward is integer RM execution within the proof-producing run. Totals are rounded independently.}
\label{tab:optimized-stages}
\begin{tabular}{@{}lrrrrrrrr@{}}
\toprule
Model/Seq & Runs & Forward & Trace & Proving & Local verify & Weight binding & Export & Total \\
\midrule
1B/256 & 3 & 1.28 & 2.45 & 6.74 & 1.40 & 37.87 & 0.04 & 49.8 \\
1B/1024 & 1 & 6.86 & 2.25 & 28.23 & 1.52 & 55.40 & 0.08 & 94.3 \\
3B/512 & 5 & 5.57 & 6.70 & 14.15 & 2.05 & 94.32 & 0.09 & 122.9 \\
8B/512 & 3 & 11.10 & 16.81 & 20.74 & 2.66 & 154.63 & 0.18 & 206.1 \\
\bottomrule
\end{tabular}
\end{table*}

\subsection{RQ3: \texorpdfstring{\brev{Public Wrapper Attestation}}{Public Wrapper Attestation}}
\begin{figure}[t]
  \centering
  \includegraphics[width=\columnwidth]{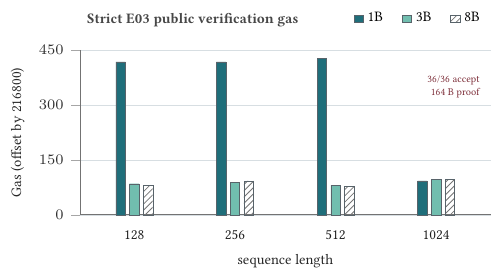}
  \caption{\brev{Public wrapper gas across the strict 12-setting public-attestation matrix.} Evidence: \textsc{R}. Gas stays near 217K across model sizes and sequence lengths because the contract checks a compact wrapper statement rather than replaying the RM proof trace.}
  \label{fig:public-gas-trend}
\end{figure}

RQ3 evaluates the optional public-attestation path. All 36 public statements (12 settings, three artifacts each) are accepted by the wrapper-generation and Hardhat verification path. The strict 12-setting matrix in Appendix Table~\ref{appx-tab:public-verification-full} records the contract-facing results: gas, proof size, and statement-binding behavior; complete composite verification is the off-chain auditor path. Fig.~\ref{fig:public-gas-trend} shows that gas stays tightly concentrated around 217K across model sizes and sequence lengths. Groth16 proof size is fixed at 164 bytes, while payload and calldata size are model-scale dependent because the exported package carries commitment and binding material.
The measured public relation is the partial wrapper in Section~\ref{sec:prototype-attestation}. Complete export and computation verification remains an off-chain auditor check.

\subsection{RQ4: Service-Bound Rejection}
RQ4 tests whether artifacts remain bound to their service context after export. We inject attacks into complete \vrmaas{} artifacts at 1B/128, 3B/512, and 8B/1024 and replay them through the off-chain artifact-verification harness. Public-field substitutions are additionally replayed through the partial wrapper and contract path. Appendix Tables~\ref{appx-tab:request-binding-tests} and~\ref{appx-tab:negative-artifacts} report the request-metadata and broader artifact tests, respectively. These full-model tests complement the small-model generic-baseline comparison in RQ6; they do not constitute a full-model throughput comparison against that baseline.

All 14 main test cases---13 attacks and one valid control---produce their expected outcomes in each of the three settings, giving 42 cases. The valid control is accepted and every attack is rejected. The attacks cover proof and reward tampering, model and weight-opening substitution, configuration and verifier changes, request replay, proof reordering, export-digest changes, and public-statement or wrapper mismatches. Four additional metadata substitutions alter the prompt digest, response digest, nonce, or policy; each is rejected through the request-binding checks. A recorded PASS means that the observed outcome matches the expected one, including acceptance of the valid control.

The rejecting path determines the guarantee: export-digest and verifier-identifier checks belong to off-chain verification, while the wrapper checks its supplied public tuple and selected consistency relations. Changing request metadata is caught by recomputing and checking the request identifier; this test does not establish a raw-text-to-hidden-state proof or an in-circuit export-digest constraint. Request binding adds 32 payload bytes, two calldata field elements, and 13,752--13,776 gas in the recorded comparisons. These measurements show how explicit acceptance fields let consumers reject otherwise transferable service artifacts at a small cost relative to producer-side proving.

\subsection{RQ5: Service Operation}
RQ5 separates the work required for each complete artifact from service-policy scenarios derived from earlier measurements. For the optimized producer, the four measured complete-run totals are 49.8--206.1 seconds and include fresh transcript-dependent weight openings. Static SRS data, checkpoint commitments, and model/configuration metadata can be reused, but the Gemini evaluation points and KZG opening proofs depend on the current request. Therefore, cache-hit or batching assumptions cannot remove this opening-generation cost from an otherwise complete artifact.

Auditors verify retained artifacts in 1.94--5.11 seconds, corresponding to approximately 704--1852 artifacts per hour for a single worker before scheduling overhead. Table~\ref{tab:optimized-verification} reports the verification-side ranges by model. For the measured 8B/512 generation setting, the producer spends 206.1 seconds, retained 8B artifacts take 4.14--5.11 seconds to audit, and partial public attestation costs about 216.9K gas. These costs support provisional service delivery followed by verification before a consequential audit or settlement action.

\begin{table}[t]
\centering
\scriptsize
\setlength{\tabcolsep}{2.4pt}
\caption{Verification-side cost by model. Values span the retained sequence lengths and artifacts. Groth16 proof size is 164 bytes and all 36 public statements are accepted. Evidence: \textsc{R}.}
\label{tab:optimized-verification}
\begin{tabular}{@{}lrrrrr@{}}
\toprule
Model & Gas & Payload B & Calldata B & Wrapper (s) & Auditor (s) \\
\midrule
1B & 216,877--217,231 & 28,580 & 11,573 & 0.09--0.10 & 1.94--2.58 \\
3B & 216,865--216,901 & 49,316 & 19,564--19,565 & 0.12--0.14 & 3.17--4.09 \\
8B & 216,889--216,901 & 56,228 & 22,229--22,230 & 0.15 & 4.14--5.11 \\
\bottomrule
\end{tabular}
\end{table}

Appendix~\ref{appx-app:additional-service-operation} preserves the original componentized amortization, batching, auditor-capacity, and queueing records. Their numerical results describe historical measurements and conditional service models. In particular, the amortization scripts charge binding generation as reusable preparation while their ``link/check'' component measures verification; that split does not isolate all per-request opening-generation work. The associated batching and queueing estimates therefore remain conditional scenarios, not measured throughput or SLA guarantees for the optimized producer. A production study must account for fresh openings, multi-request scheduling, and cache invalidation explicitly.

\subsection{RQ6: Design Necessity and Generic Baseline}
RQ6 asks whether service binding adds guarantees beyond a generic inference proof. Service bindings define the accepted statement, while witness construction and opening aggregation reduce producer work. For the optimized implementation, computation proving averages 14.3 seconds over 1B/512, 3B/512, and 8B/512; a specified analytical variant without matmul aggregation estimates 19.4 seconds ($1.35\times$), rather than a separately executed prover measurement. The original ablation rows remain in Appendix Table~\ref{appx-tab:component-ablation} as historical records; their cache-based P4 totals depend on the conditional amortization model described in Appendix~\ref{appx-app:additional-service-operation}.

We run GenericProof+Wrapper and a reduced \vrmaas{} variant retaining all service bindings on two small models: a one-layer RM and a two-layer Transformer. Both models yield the same attack outcomes, reported in Appendix Table~\ref{appx-tab:generic-attack-outcomes}. This comparison isolates verifier-facing statement semantics. The comparison focuses on whether service fields are first-class accepted statement fields, rather than on prover throughput. This is the key service distinction: a proof may certify that some computation was performed, yet still leave the request, deployment configuration, verifier identity, or export digest outside the object that downstream systems accept. Downstream consumers can rely on those fields only when an acceptance check binds them to the intended service context. GenericProof+Wrapper blocks proof-byte and output tampering, but accepts requestID, configID, model commitment, verifierID, proof-order, and stale-replay substitutions because those fields are outside its statement. The reduced \vrmaas{} variant rejects all eight attack classes. \rev{This result clarifies the novelty boundary: \vrmaas{} builds on standard proof components, but changes the accepted statement from a raw inference proof to a request-, model-, configuration-, verifier-, order-, and export-bound reward-service output.}

\subsection{RQ7: Configuration and Validity Limits}
RQ7 records configuration limits that a deployer should expose rather than hide. The current-configuration diagnostic in Appendix Table~\ref{appx-tab:quant-config-tradeoff} shows how deployments can classify verifier configurations as registered, registered with warning, or flagged for finer quantization. Pairwise agreement remains 93.5--97.5\% for the retained long-context rows, but the 8B/1024 setting reaches mean absolute error 2.326 and P95 absolute error 7.332, so it is flagged for finer configuration.

A deployment can expose this decision as a config-level admissibility policy. In our prototype, a configuration is marked \textsc{Accept} when $\mathrm{PairAgree}\geq 0.97$, $\mathrm{MAE}\leq 0.50$, $\mathrm{P95}\leq 2.00$, and $R^2\geq 0.95$; \textsc{Warn} when $\mathrm{PairAgree}\geq 0.935$, $\mathrm{MAE}\leq 2.00$, and $\mathrm{P95}\leq 5.00$; and \textsc{Retune} otherwise. These thresholds are deployment parameters, not proof-system assumptions: the proof guarantees execution under the declared numerical semantics, while the service policy decides whether that semantics is admissible.

\section{Discussion and Threats to Validity}
\label{sec:discussion-threats}

\noindent\hspace*{\parindent}\textbf{Deployment scope.}
Computation proving and weight binding dominate producer-side latency, so the design targets delayed audit, settlement, or governance rather than low-latency serving.

\noindent\hspace*{\parindent}\textbf{Service and semantic scope.}
\vrmaas{} certifies faithful execution of a declared reward service: an accepted artifact follows the declared RM computation, committed weights, verifier implementation, numerical configuration, and request context. It does not certify that the RM is fair, unbiased, robust to reward hacking, or aligned with every governance objective; those concerns remain part of model and policy governance. Long-context quantization drift makes $\mathsf{configID}$, lookup-table versions, scale factors, and reward-export rules explicit service-configuration choices.

\noindent\hspace*{\parindent}\textbf{\brev{Privacy scope.}}
\brev{The public statement does not publish the full checkpoint, hidden activations, or weight-opening transcript. Off-chain weight binding still exposes opening values such as $v_j$ to auditors; stronger auditor-side weight hiding is an orthogonal commitment-opening design choice.}

\noindent\hspace*{\parindent}\textbf{Input and attestation boundaries.}
The computation proof begins with field-encoded hidden states supplied in the canonical RM-input package. Embedding lookup and the derivation of input padding/segment semantics and the last-token position from the prompt/response digest are not proved by the current implementation. The optimized trace includes attention-score and value-aggregation products, but request-metadata binding alone does not establish this raw-text-to-hidden-state link. Similarly, the measured public wrapper checks the partial relation in Section~\ref{sec:prototype-attestation}; the complete computation and export guarantees require off-chain audit.

\noindent\hspace*{\parindent}\textbf{Evaluation scope.}
The experiments use a high-performance single-node setting. The original batching and queueing studies replay historical stage latencies under a componentized caching assumption; they have not been rerun as a production deployment of the optimized producer. The optimized latency rows contain one to five completed runs per setting. Their measurements establish the reported per-request costs, while multi-tenant scheduling, admission control, cache invalidation, checkpoint rotation, and SLA-aware verifier placement remain future work. Caching static checkpoint data, true tensor batching, and workflow-specific $\mathsf{configID}$ admission are further engineering steps, with transcript-dependent openings regenerated for each request.

\noindent\hspace*{\parindent}\textbf{Beyond reward models.}
\rev{A moderation label, an LLM-as-judge score, or a ranking decision raises the same service-binding questions. The artifact fields, split verification, and lifecycle interfaces can apply to these outputs, while their computation proofs must reflect the served model and its output semantics. For the RM studied here, the last-token reward head and its task chain define that output-specific part of the proof. Extending the interface requires adapting the computation relation and measuring its producer cost.}

\noindent\hspace*{\parindent}\rev{\textbf{Further producer optimization.}
Weight binding remains the dominant cost after request-level opening batching. Further work can target its GPU kernels and memory traffic, together with witness construction and lookup-workspace reuse. The integer witness path requires accumulation bounds that fit within signed 64-bit integers; other numerical configurations require checking those bounds.}

\section{Related Work}
\label{sec:related}

\noindent\hspace*{\parindent}\textbf{Reward-model lifecycle.}
Reward-model work studies how to train, evaluate, and diagnose preference models for RLHF and alignment workflows~\cite{stiennon2020learning,ouyang2022training,malik2025rewardbench2advancingreward,liu2025skywork,perez2023discovering,bu2025beyond,shen2023loose}. Verifiable-reward methods make outcome signals externally checkable~\cite{liu2025trust}. \vrmaas{} addresses the deployed-service stage: once an RM is selected, auditors need an accepted statement binding each returned score to the request, checkpoint, numerical configuration, verifier implementation, and export semantics.

\noindent\hspace*{\parindent}\textbf{Verifiable computation and ZKML.}
\rev{General-purpose proof systems~\cite{parno2016pinocchio,xie2019libra,setty2020spartan,groth2016,gabizon2019plonk,chiesa2020marlin} and research on verifiable or zero-knowledge ML~\cite{ghodsi2017safetynets,liu2021zkcnn,zhao2021veriml,huang2022zkmlaas,peng2026survey,sun2024zkllm,qu2025zkgpt} enable proofs of computation on untrusted infrastructure, with zkLLM-style systems reaching billion-parameter Transformers. Specialized protocols address convolutional networks~\cite{liu2021zkcnn}, training and fine-tuning~\cite{sun2024zkdl,liao2026verilora}, and inference compilation and scaling~\cite{feng2021zen,kang2022scaling,weng2021mystique}. \vrmaas{} builds on their sumcheck, lookup, and commitment toolbox~\cite{thaler2022proofs,campanelli2024lookup} to bind an RM computation to its service context. Appendix Table~\ref{appx-tab:related-comparison} summarizes the fields represented in the cited formulations: computation statements certify specified operations and, in some systems, committed weights, while a deployed service must also define how requests, configurations, verifiers, task order, and exports affect acceptance. Generic proof systems can incorporate these bindings; \vrmaas{} specifies them for RM services and measures their enforcement cost. Appendix Table~\ref{appx-tab:generic-attack-outcomes} reports the resulting distinction in our evaluation: the GenericProof+Wrapper baseline accepts six service-field substitutions that \vrmaas{} rejects.}

\noindent\hspace*{\parindent}\textbf{Trusted hardware and optimistic verification.}
\rev{TEE-based attestation~\cite{costan2016intel} and TEE-assisted inference such as Slalom~\cite{tramer2019slalom} rely on the hardware vendor and attested enclave execution; Slalom checks outsourced linear layers inside the enclave. Optimistic ML systems such as opML~\cite{conway2024opml} resolve disputes through fraud proofs within a challenge window, requiring an honest challenger with access to the model and other data needed for re-execution. \vrmaas{} supports delayed settlement through proofs under the stated cryptographic assumptions: anyone can check an artifact without rerunning inference or obtaining the private checkpoint, at the producer cost measured in RQ2. These approaches can compose, with rewards accepted optimistically and a \vrmaas{} artifact required upon dispute.}

\noindent\hspace*{\parindent}\textbf{Service computing and attestation.}
\rev{Service-oriented computing studies cross-organization composition, dependable delivery, and integrity-aware interfaces~\cite{huhns2005service,papazoglou2007service}, while privacy-preserving distributed analytics verifies results across mutually distrustful parties~\cite{froelicher2020drynx}. ML-as-a-service verification~\cite{zhao2021veriml,huang2022zkmlaas} brings proofs to outsourced inference. \vrmaas{} connects these perspectives through a reward artifact whose acceptance rule governs delivery, audit, and settlement. Its lifecycle interfaces let organizations retain and reuse evidence, with complete off-chain audit and partial public attestation serving different consumers.}

\section{Conclusion}
\label{sec:conclusion}

This paper reformulates reward-model inference as a services-computing problem: an RM service should return a request-bound artifact, not only a score. \vrmaas{} realizes this idea through proof-carrying reward outputs, request and model-weight binding, verifier/configuration identifiers, export semantics, and optional public attestation. By checking these bindings at acceptance, the same RM result can travel across requesters, auditors, and settlement logic with reusable evidence. The prototype preserves 93.5--99\% pairwise agreement with the baseline RM and rejects the six service-field substitution classes accepted by the GenericProof+Wrapper baseline. Complete off-chain verification and partial public attestation provide distinct checks for these consumers. Exact integer witness construction and request-level opening aggregation reduce complete artifact generation to 49.8--206.1 seconds across the evaluated latency settings, with auditor verification in 1.94--5.11 seconds and partial public attestation near 217K gas. Weight binding remains the dominant producer component, and transcript-dependent openings are generated for every request. These measurements support delayed audit, settlement, and governance. The practical service question is therefore how providers schedule proof work around audit and settlement windows while keeping verifier-facing outputs portable. The broader direction is to make LLM-alignment service outputs portable and auditable without forcing requesters to trust black-box AI APIs.

\clearpage
\appendices
\section*{Appendix Overview}

These appendices provide the protocol background and extended experimental records that support the main text. They are organized to keep the main paper focused on the service abstraction, artifact semantics, and primary results, while still making the cryptographic relations, measurement context, and diagnostic evidence inspectable.

\textbf{Appendix roadmap:}
Appendix~\ref{appx-app:proof-components} records the proof-system components used by the \vrmaas{} prototype: polynomial commitments, Gemini folding, Groth16 wrapping, sumcheck-style arithmetic checks, tensor relations, and lookup arguments. Appendix~\ref{appx-app:evaluation-matrix} summarizes the experimental axes. Appendix~\ref{appx-app:additional-fidelity} separates benchmark-label accuracy from service fidelity, and Appendix~\ref{appx-app:additional-latency} reports the no-proof baseline latency used to contextualize proof-carrying cost. Appendix~\ref{appx-app:additional-public-verification} expands the \brev{public-attestation matrix} behind the gas trend in the main paper. Appendix~\ref{appx-app:additional-rejection} gives the request-binding and negative-artifact rejection matrices. Appendix~\ref{appx-app:additional-service-operation} provides the service-operation evidence for batching, auditor capacity, and trace-driven queueing. Appendices~\ref{appx-app:additional-ablation}, \ref{appx-app:additional-generic-baseline}, and~\ref{appx-app:additional-config} give the component-ablation, generic-baseline, and configuration-drift diagnostics used by RQ6 and RQ7. Appendix~\ref{appx-app:case-level-fidelity} closes with compact case-level examples that illustrate why \vrmaas{} treats fidelity as equivalence to the declared RM service rather than as a claim about reward-model semantic quality. Appendix~\ref{appx-app:security-mapping} records the detailed threat-to-guarantee mapping, and Appendix~\ref{appx-app:related-work} gives the extended related-work comparison. Appendix~\ref{appx-app:historical-latency} preserves producer latencies before optimization, and Appendix~\ref{appx-app:extended-attestation} specifies the conditional export-binding public relation.

\section{Cryptographic Components and Proof Relations}
\label{appx-app:proof-components}

This section records the protocol-level background used by the \vrmaas{} prototype. The main body uses these components to define service-facing proof tasks, model-weight binding, and public attestation. Here we spell out the corresponding algebraic relations and cryptographic interfaces. The \vrmaas{} framework does not claim new primitives; it composes standard proof, lookup, commitment, and succinct-wrapper tools around reward-model serving.

\subsection{Proofs, Arguments, and Polynomial Commitments}

\subsubsection{Proofs and Arguments}
Interactive verification protocols are commonly separated into \emph{proofs} and \emph{arguments}. A proof provides information-theoretic soundness, while an argument provides computational soundness against polynomial-time adversaries. In both cases, the protocol is expected to satisfy three properties:
\begin{itemize}
    \item \textbf{Completeness}: if the statement is true and the prover follows the protocol, the verifier accepts except with negligible probability.
    \item \textbf{Soundness}: if the statement is false, no malicious prover can make the verifier accept except with negligible probability.
    \item \textbf{Zero knowledge}: the verifier learns nothing beyond the truth of the statement.
\end{itemize}
Here $\lambda$ denotes the security parameter and $\mathsf{negl}(\lambda)$ denotes a negligible function.

\begin{definition}[Interactive proof system]
\label{appx-def:app-ip}
Let $L \subseteq \{0,1\}^{*}$ be an NP language with witness relation $R=\{(x,w)\}$. An interactive proof system is a pair of probabilistic polynomial-time algorithms $(\mathcal{P},\mathcal{V})$ that execute an interactive protocol on common input $x$, where $\mathcal{P}$ additionally holds private witness $w$. The system is complete if, for $x \in L$ and $(x,w)\in R$,
\[
\Pr[\langle \mathcal{P}(x,w),\mathcal{V}(x)\rangle=1]
\geq 1-\mathsf{negl}(\lambda),
\]
and sound if, for $x \notin L$ and any PPT prover $\mathcal{P}^{*}$,
\[
\Pr[\langle \mathcal{P}^{*}(x),\mathcal{V}(x)\rangle=1]
\leq \mathsf{negl}(\lambda).
\]
\end{definition}

Public-coin proof systems derive verifier messages from public randomness. SNARK systems build on this model to provide short proofs and efficient verification~\cite{gabizon2019plonk,chiesa2020marlin,parno2016pinocchio}.

\begin{definition}[Zero-knowledge proof]
\label{appx-def:app-zk}
An interactive proof system is zero knowledge if there exists a PPT simulator $\mathcal{S}$ such that, for every instance $(x,w)\in R$ and every PPT distinguisher $\mathcal{D}$,
\[
\left|
\Pr[\mathcal{D}(\mathsf{Real}_{\mathcal{P},\mathcal{V}}(x,w))=1]
-
\Pr[\mathcal{D}(\mathcal{S}(x))=1]
\right|
\leq
\mathsf{negl}(\lambda),
\]
where $\mathsf{Real}_{\mathcal{P},\mathcal{V}}(x,w)$ denotes the verifier's view in the real protocol execution.
\end{definition}

\subsubsection{Polynomial Commitments}
\label{appx-def:app-polycommit}
A polynomial commitment scheme lets a prover commit to a polynomial and later prove evaluations at verifier-chosen points. In the \vrmaas{} framework, this role is instantiated by KZG-style commitments~\cite{kate2010constant}, which provide constant-size commitments and succinct evaluation proofs.

We work over a bilinear group $(\mathbb{G}_1,\mathbb{G}_2,\mathbb{G}_T)$ of prime order $|\mathbb{F}|$ with generators $g_1\in\mathbb{G}_1$ and $g_2\in\mathbb{G}_2$ and a non-degenerate pairing $e:\mathbb{G}_1\times\mathbb{G}_2\rightarrow\mathbb{G}_T$. For a scalar $x\in\mathbb{F}$ we write $[x]_1=x\,g_1$ and $[x]_2=x\,g_2$ for its encodings in $\mathbb{G}_1$ and $\mathbb{G}_2$; in particular $[1]_1=g_1$ and $[1]_2=g_2$. The prototype instantiates this group with the BN254 curve.

Let $f(X)=\sum_{i=0}^{D_{\mathrm{KZG}}} a_i X^i \in \mathbb{F}[X]$. For a secret trapdoor $\tau$, the structured reference string is
\begin{equation}
\mathsf{SRS}=\left([1]_{1},[\tau]_{1},\ldots,[\tau^{D_{\mathrm{KZG}}}]_{1},[1]_{2},[\tau]_{2}\right).
\label{appx-eq:app-kzg-srs}
\end{equation}
The commitment is
\begin{equation}
C=[f(\tau)]_{1}=\sum_{i=0}^{D_{\mathrm{KZG}}} a_i[\tau^i]_{1}.
\label{appx-eq:app-kzg-commit}
\end{equation}
For an evaluation point $z\in\mathbb{F}$ and value $y=f(z)$, the prover forms
\begin{equation}
q(X)=\frac{f(X)-y}{X-z}
\label{appx-eq:app-kzg-quotient}
\end{equation}
and outputs $\pi=[q(\tau)]_{1}$. The verifier accepts iff
\begin{equation}
e(C-y[1]_{1},[1]_{2}) = e(\pi,[\tau]_{2}-z[1]_{2}).
\label{appx-eq:app-kzg-verify}
\end{equation}
In the \vrmaas{} framework, polynomial commitments authenticate proof-relevant tensor and weight evaluations and support the model-weight binding material defined in the main text.

\subsubsection{Gemini Folding for Multilinear Openings}
\label{appx-app:gemini}
Many \vrmaas{} proof claims arise from multilinear extensions inside sumcheck-style verification. Rather than using a separate multilinear commitment interface, the implementation uses Gemini folding~\cite{bootle2022gemini} to reduce multilinear opening claims to a compact set of univariate openings discharged by the KZG backend. Gemini recursively folds a multilinear polynomial with transcript-derived challenges, preserving the claimed evaluation while reducing the number of variables round by round. Thus, Gemini bridges MLE-based tensor relations and the univariate polynomial-commitment layer.

\subsubsection{Groth16}
\label{appx-app:groth16}
Groth16~\cite{groth2016} is a preprocessing zkSNARK for arithmetic circuits and rank-1 constraint systems. Given an R1CS instance
\begin{equation}
(Az)\circ(Bz)=Cz,
\label{appx-eq:app-groth16-r1cs}
\end{equation}
where $\circ$ is the entrywise (Hadamard) product, $A,B,C$ are the public constraint matrices, and the assignment vector $z=(1,x,w)$ stacks the constant $1$, the public inputs $x$, and the witness $w$; setup produces a proving key $\mathsf{pk}$ and verifying key $\mathsf{vk}$. The prover outputs
\begin{equation}
\pi=(\pi_A,\pi_B,\pi_C)\in \mathbb{G}_1\times \mathbb{G}_2\times \mathbb{G}_1,
\label{appx-eq:app-groth16-proof}
\end{equation}
and the verifier checks a constant number of pairing equations.

In the \vrmaas{} framework, Groth16 is the final public wrapper. \brev{The design uses split verification: the complete exported verifier runs off chain, while the Groth16 circuit proves a compact attestation predicate for $\mathrm{Stmt}^{\mathrm{proto}}_{\mathrm{pub}}$.} The implemented wrapper keeps selected public-field, reward-binding, and fold/sumcheck-consistency checks contract-consumable while the chain sees only the public statement and $\pi_G$. Thus, Groth16 provides the deployed succinctness and zero-knowledge boundary, whereas the inner arithmetic, lookup, and commitment checks establish service-output integrity \brev{through the off-chain verification path}.

\subsubsection{Sumcheck Protocol}
\label{appx-app:sumcheck}
The sumcheck protocol~\cite{lund1992algebraic,cormode2012practical} verifies multivariate polynomial sums. For a polynomial $f:\mathbb{F}^{\nu}\rightarrow\mathbb{F}$ of individual degree at most $\delta$, the prover claims
\begin{equation}
S=\sum_{b_1,\ldots,b_{\nu}\in\{0,1\}} f(b_1,\ldots,b_{\nu}).
\label{appx-eq:app-sumcheck}
\end{equation}
The protocol runs for $\nu$ rounds. At each round, the prover sends a univariate polynomial encoding a partial sum and the verifier samples a random challenge $r_i\in\mathbb{F}$. The final claim reduces to a single evaluation of $f$. By Schwartz--Zippel~\cite{schwartz1980fast,zippel1979probabilistic}, the soundness error is bounded by $\delta/|\mathbb{F}|$ for each polynomial identity check. In the \vrmaas{} framework, sumcheck-style arguments are used for arithmetic tensor tasks such as matrix multiplication, residual addition, and element-wise products.

\subsubsection{Multilinear Extensions}
\label{appx-app:mle}
Multilinear extensions (MLEs) lift Boolean functions to finite-field polynomials. For $f:\{0,1\}^{\nu}\rightarrow\mathbb{F}$, the unique multilinear extension $\widetilde{f}:\mathbb{F}^{\nu}\rightarrow\mathbb{F}$ is
\begin{equation}
\widetilde{f}(\mathbf{x})
=
\sum_{\mathbf{b}\in\{0,1\}^{\nu}}
f(\mathbf{b})
\prod_{i=1}^{\nu}\big(b_i x_i+(1-b_i)(1-x_i)\big).
\label{appx-eq:app-mle}
\end{equation}
The product term is the multilinear Lagrange basis: it evaluates to $1$ when $\mathbf{x}=\mathbf{b}$ and to $0$ on other Boolean points. This representation lets tensor entries be embedded into low-degree polynomial relations and is a standard basis for efficient sumcheck verification~\cite{setty2020spartan}.

\subsubsection{Fiat--Shamir Transform and Transcript Binding}
\label{appx-app:fiat-shamir}
The arithmetic sumchecks, lookup arguments, and commitment openings above are public-coin interactive protocols: every verifier message is a uniform random challenge. The \vrmaas{} prototype makes them non-interactive through the Fiat--Shamir transform, deriving each challenge from a cryptographic hash of the running transcript, instantiated with the BN254-native Poseidon2 sponge. Soundness of this transform is not automatic. The \emph{strong} Fiat--Shamir transform requires that, before any challenge is derived, the transcript absorb the full public statement together with all prior prover messages and commitments; deriving a challenge over an incomplete transcript admits the weak Fiat--Shamir attacks of Bernhard et al.~\cite{bernhard2012not}, in which a malicious prover fixes a commitment or a public input only after seeing the challenge. Accordingly, the prototype absorbs, in the fixed order $\mathsf{op\_order}$ and under domain-separated tags, every ordered task descriptor, tensor and weight commitment, and prover proof message, as well as the public statement fields bound to the artifact---the request identifier, model-weight commitment, verifier identifier, and configuration identifier---before each challenge is drawn. This binds the derived challenges to a single trace and to one declared service configuration, so a challenge cannot be reused across a different wiring, a different lookup-table version, or a different request. Modeling Poseidon2 as a domain-separated random oracle, a prover making at most $Q$ transcript queries forfeits at most $Q/2^{\kappa}$ in soundness, where $\kappa$ is the transcript output length in bits.

\subsection{Verifiable Tensor Operations}
\label{appx-app:tensor-ops}
Reward-model inference is dominated by matrix and tensor operations. Let $A\in\mathbb{F}^{D_0\times D_1}$, $B\in\mathbb{F}^{D_1\times D_2}$, and $C\in\mathbb{F}^{D_0\times D_2}$ satisfy $C=A\cdot B$. Element-wise, this means
\begin{equation}
C[i,j] = \sum_{k=1}^{D_1} A[i,k]B[k,j],
\qquad
i<D_0,\ j<D_2.
\label{appx-eq:app-matmul}
\end{equation}
Assume, after zero padding if needed, that $D_0=2^{d_0}$, $D_1=2^{d_1}$, and $D_2=2^{d_2}$. Lifting $A$, $B$, and $C$ to MLEs $\widetilde{A}$, $\widetilde{B}$, and $\widetilde{C}$ yields the matrix-multiplication identity
\begin{equation}
\widetilde{C}(u,v)
=
\sum_{k\in\{0,1\}^{d_1}}
\widetilde{A}(u,k)\,\widetilde{B}(k,v).
\label{appx-eq:app-zk-mle-matmul}
\end{equation}
Here $u$ and $v$ are \emph{formal field variables}, not Boolean indices: both sides of~\eqref{appx-eq:app-zk-mle-matmul} are multilinear in $(u,v)$ and agree on the Boolean hypercube $\{0,1\}^{d_0}\times\{0,1\}^{d_2}$, where they reduce to the entrywise product~\eqref{appx-eq:app-matmul}, so by uniqueness of the multilinear extension they coincide for all $u\in\mathbb{F}^{d_0}$ and $v\in\mathbb{F}^{d_2}$. Soundness comes from instantiating $(u,v)$ at a random point $(r_0,r_2)\in\mathbb{F}^{d_0}\times\mathbb{F}^{d_2}$: the verifier runs a $d_1$-round sumcheck on the right-hand sum over $k$ against the claimed value $\widetilde{C}(r_0,r_2)$, the sumcheck reduces the contraction index $k$ to a random point $r_1\in\mathbb{F}^{d_1}$, and the final claim is discharged by opening the committed MLEs at $\widetilde{A}(r_0,r_1)$ and $\widetilde{B}(r_1,r_2)$. For a single task the verifier samples $(r_0,r_2)$ directly; in the batched setting of the main text, all same-shaped tasks share one aggregation challenge, so every task in a batch is checked at the same $(r_0,r_2)$. This is the algebraic basis for the matrix-multiplication proof tasks and their aggregation strategy in the main text.

\subsection{Lookup Arguments for Non-Arithmetic Operations}
\label{appx-app:lookup}
Transformer reward models also contain table-defined and non-arithmetic behavior: softmax-style normalization, discretized normalization maps, activation maps, range checks, and fixed-point rescaling. Lookup arguments prove that private values or input-output pairs belong to configured public tables~\cite{thaler2022proofs,kang2022scaling,sun2024zkllm}.

Let $S=\{s_1,\ldots,s_D\}\subseteq\mathbb{F}$ be a private witness multiset and $T=\{t_1,\ldots,t_N\}\subseteq\mathbb{F}$ be a public table. The raw witness size depends on the tensor shape of the proof task that feeds the table, so before running the argument the prover pads $S$ to the next multiple of the table size $N$ with a designated table element $t_1\in T$; we let $D$ denote this padded size, so that $N\mid D$. Because the padding values already belong to $T$, padding only raises the multiplicity $m_1$ of $t_1$ and leaves the membership relation $S\subseteq T$, and hence the soundness of the argument, unchanged. To prove $S\subseteq T$, the prover defines multiplicities $m=\{m_1,\ldots,m_N\}$ and proves the rational identity
\begin{align}
\frac{d}{dX}\log\!\left(\prod_{i=1}^{D}(X+s_i)\right)
=
\frac{d}{dX}\log\!\left(\prod_{j=1}^{N}(X+t_j)^{m_j}\right).
\label{appx-eq:app-lookup-rational}
\end{align}
Equivalently, at verifier challenge $\beta\in\mathbb{F}$, define
$A_i=(\beta+s_i)^{-1}$ and $B_j=(\beta+t_j)^{-1}$. The membership claim becomes
\begin{equation}
\sum_{i\in[D]} A_i
=
\sum_{j\in[N]} m_jB_j.
\label{appx-eq:app-lookup-inner}
\end{equation}
To make~\eqref{appx-eq:app-lookup-inner} checkable by a single sumcheck, the prover commits to the inverse vectors $A$ and $B$ and proves their well-formedness, $A_i(\beta+s_i)=1$ for $i\in[D]$ and $B_j(\beta+t_j)=1$ for $j\in[N]$, together with the inner-product equality~\eqref{appx-eq:app-lookup-inner}. Batching these claims with a verifier challenge $\alpha$ and the multilinear equality basis $\widetilde{e}$ of~\eqref{appx-eq:app-mle} reduces them to a single sumcheck instance, following the tlookup construction of zkLLM~\cite{sun2024zkllm}. In \vrmaas{}, Protocol~\ref{appx-alg:lookup-argument} instantiates this interface for config-bound mapping and range lookups.

\begin{algorithm}[t]
\floatname{algorithm}{Protocol}
\caption{\textsc{Lookup Argument}}
\label{appx-alg:lookup-argument}
\begin{algorithmic}[1]
\Require Prover holds private multiset $S\in\mathbb{F}^{D}$.
\Require Verifier fixes the configured public table $T\in\mathbb{F}^{N}$, with $N\mid D$ ensured by padding $S$ with a table element.
\State \textbf{Setup:} bind the table by $\mathsf{cm}_{T}\gets\mathsf{Commit}(T;0)$ under the declared $\mathsf{configID}$.
\State \textbf{Preparation:} prover computes the claimed multiplicity vector $m=\mathsf{m}(S,T)$.
\State Prover sends $\mathsf{cm}_{S}\gets\mathsf{Commit}(S)$ and $\mathsf{cm}_{m}\gets\mathsf{Commit}(m)$.
\State Verifier samples $\beta\leftarrow\mathbb{F}$; a zero denominator is treated as an abort event.
\State Prover computes $A_i=(\beta+s_i)^{-1}$ and $B_j=(\beta+t_j)^{-1}$.
\State Prover sends $\mathsf{cm}_{A}\gets\mathsf{Commit}(A)$ and $\mathsf{cm}_{B}\gets\mathsf{Commit}(B)$.
\State Verifier samples $\alpha\leftarrow\mathbb{F}$.
\State Prover and verifier run the batched sumcheck for inverse well-formedness and the membership equality~\eqref{appx-eq:app-lookup-inner}.
\State Prover opens the committed evaluations required at the final sumcheck point.
\State Verifier accepts iff the sumcheck, table binding, and all commitment openings verify.
\end{algorithmic}
\end{algorithm}

\subsection{Compositional Soundness of the Proof Trace}
\label{appx-app:composition}

The previous subsections establish soundness for each task family in isolation: sumcheck for arithmetic tensor relations, lookup arguments for table-defined and range claims, and polynomial-commitment openings for the committed operands. A reward-model artifact, however, asserts a single statement about an end-to-end forward pass, so we must argue that the $m$ individual task proofs compose into one consistent claim about the prescribed RM execution. Two ingredients are needed: \emph{wiring consistency}, which forces the output of one task to be the input of the next, and a \emph{union bound} over all task-level soundness errors. Throughout, $\delta$ bounds the individual degree of every sumcheck polynomial, $\ell_{\max}$ bounds the number of variables of any committed tensor handle (equivalently, the rounds of its sumcheck or Gemini folding), and $|\mathbb{F}|\approx 2^{254}$ is the prototype's BN254 scalar field.

\begin{lemma}[Trace wiring consistency]
\label{appx-lem:app-wiring}
Let $(\tau_1,\ldots,\tau_m)$ be the deterministic proof trace of the main text, where each descriptor $\tau_i$ exposes input and output handles $\mathsf{in}_i,\mathsf{out}_i$ that reference committed or witness-side tensors. Suppose the composite proof is accepted, so that (i) every task relation $\mathcal{R}_{\tau_i}=1$; (ii) for each data dependency $\mathsf{out}_i\to\mathsf{in}_{i'}$ the corresponding \textsf{equality} task certifies that the two handles open to the same committed value; and (iii) every proof-relevant weight handle satisfies the weight-binding link of the main text against $\mathcal{C}_W$. Then, except with probability at most $c_w\,m\,\ell_{\max}\,\delta/|\mathbb{F}|$ over the transcript challenges---where $c_w$ bounds the number of input handles consumed by a single task---there is a single assignment to the intermediate states $\{H^{(\ell)}\}$ that simultaneously satisfies all $m$ tasks under the declared $(M,W,\mathsf{configID})$.
\end{lemma}

\begin{proof}[Proof sketch]
Each \textsf{equality} task is a multilinear identity between two committed handles over at most $\ell_{\max}$ variables, checked at a transcript-derived random point; by the Schwartz--Zippel bound used in Appendix~\ref{appx-app:sumcheck}, a single such check fails to detect a mismatch with probability at most $\ell_{\max}\,\delta/|\mathbb{F}|$. Since each task consumes a bounded number of input handles, the trace has $O(m)$ data-dependency edges, and a union bound over them gives $c_w\,m\,\ell_{\max}\,\delta/|\mathbb{F}|$. Because the ordered descriptors, commitments, and proof messages are absorbed into the Poseidon2 transcript in the fixed order $\mathsf{op\_order}$, the challenges are bound to one trace and cannot be reused to splice handles from a different wiring. Finally, the weight-binding link ties each opened weight operand to $\mathcal{C}_W$, so the consistent assignment uses the registered checkpoint rather than free witness values.
\end{proof}

\begin{theorem}[Composite soundness]
\label{appx-thm:app-composite}
Let the trace contain $m$ tasks and let the matrix-multiplication aggregation combine at most $B$ same-shaped tasks per batch, with $\delta$ and $\ell_{\max}$ as defined above. Model the Poseidon2 transcript as a random oracle queried at most $Q$ times, assume the polynomial-commitment scheme is binding with advantage $\epsilon_{\mathrm{bind}}$, and assume the lookup tables are fixed by $\mathsf{configID}$. Then any probabilistic polynomial-time prover that makes the off-chain verifier accept a trace that deviates from the prescribed RM execution succeeds with probability at most
\begin{equation}
\epsilon_{\mathrm{comp}}
\le
\frac{c\,(m\,\ell_{\max}\,\delta + B)}{|\mathbb{F}|}
+ \epsilon_{\mathrm{bind}}
+ \frac{Q}{2^{\kappa}},
\label{appx-eq:app-composite-union-bound}
\end{equation}
where $c$ is a small constant counting the arithmetic, lookup, equality-wiring, and folding checks per task and $\kappa$ is the transcript output length in bits. For the prototype's parameters ($m\le 2^{15}$ tasks, $\ell_{\max}\le 24$, $\delta\le 3$, $B\le 2^{10}$, $\kappa=254$), the algebraic field-checking term stays below $2^{-220}$. \brev{This bounds only the statistical (Schwartz--Zippel) error of the interactive checks; the overall security of the deployed system is capped by the computational security of the BN254-based commitment and SNARK assumptions ($\epsilon_{\mathrm{bind}}$) and by the random-oracle term $Q/2^{\kappa}$. We therefore do not claim a $200$-bit \emph{system} security level, only that the algebraic field-checking error is far from being the binding term.}
\end{theorem}

\begin{proof}[Proof sketch]
We bound the probability that the verifier accepts while some task deviates from the prescribed execution. By Lemma~\ref{appx-lem:app-wiring}, the accepted handles are wired into a single assignment except with probability $c_w\,m\,\ell_{\max}\,\delta/|\mathbb{F}|$. Conditioned on consistent wiring, a deviating execution must falsify at least one task relation, so it suffices to sum the per-check soundness errors. Each arithmetic sumcheck contributes at most $\nu_i\,\delta/|\mathbb{F}|$ over its rounds; the matrix-multiplication aggregation adds at most $(B-1)/|\mathbb{F}|$ per batch from the random linear combination~\eqref{appx-eq:app-zk-mle-matmul}; each lookup argument contributes a sumcheck error of the same order through Protocol~\ref{appx-alg:lookup-argument}; and each multilinear opening discharged by Gemini folding contributes at most $\nu_i\,\delta/|\mathbb{F}|$ across its folding rounds before the final commitment check, whose binding failure is absorbed into $\epsilon_{\mathrm{bind}}$. Collecting the $O(m)$ checks, each with at most $\ell_{\max}$ rounds and individual degree at most $\delta$, yields the field term $c\,(m\,\ell_{\max}\,\delta + B)/|\mathbb{F}|$. The term $Q/2^{\kappa}$ bounds the advantage of a random-oracle prover making $Q$ transcript queries under the strong Fiat--Shamir transform of Appendix~\ref{appx-app:fiat-shamir}. Substituting the prototype parameters gives the stated bit-security.
\end{proof}

Thus an accepted composite proof certifies not a collection of unrelated task proofs but a single wired execution whose final reward-head output is $r_{\mathrm{pub}}$, and the configuration identifier fixes the lookup tables and quantization rules under which this execution is interpreted.

\section{Evaluation Matrix}
\label{appx-app:evaluation-matrix}

Table~\ref{appx-tab:eval-matrix} summarizes the models, sequence lengths, workload, and evidence types covered by the \vrmaas{} evaluation.

\begin{table}[!t]
  \centering
  \small
  \caption{Evaluation matrix for the \vrmaas{} prototype.}
  \label{appx-tab:eval-matrix}
  \begin{tabular}{p{0.30\columnwidth}p{0.60\columnwidth}}
    \toprule
    \textbf{Axis} & \textbf{Configuration} \\
    \midrule
    Models & Skywork reward models at 1B, 3B, and 8B scales~\cite{liu2025skywork} \\
    Sequence lengths & 128, 256, 512, and 1024 \\
    Workload & Repeated QA comparison reward scoring; E02 reports 200 chosen/rejected cases per model/sequence setting \\
    Direct measurements & Fidelity, no-proof baseline latency, proof-stage latency, wrapper verification, request-bound contract tests, negative-artifact tests \\
    Derived or replay evidence & P4 amortization, service-window batching replay, auditor workload replay, trace-driven queueing simulation, specified ablation variants \\
    Deployment metrics & Gas, payload size, Groth16 proof size, calldata size, invalid-artifact rejection, request/config/model/verifier binding, queueing delay \\
    \bottomrule
  \end{tabular}
\end{table}

\section{Extended Fidelity Diagnostics}
\label{appx-app:additional-fidelity}

\begin{figure}[!t]
	\centering
	\includegraphics[width=\columnwidth]{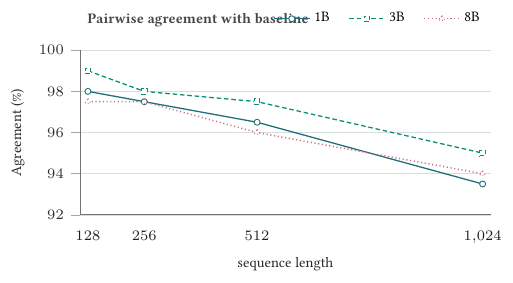}
	\caption{Pairwise service-fidelity trend across model sizes and context lengths. Pairwise agreement compares the proof-aware preference with the baseline RM-service preference, not with the benchmark label.}
	\label{appx-fig:pairwise-agreement-trend}
\end{figure}

The main manuscript reports service fidelity to the baseline reward-model path. Table~\ref{appx-tab:benchmark-context} separately reports benchmark preference accuracy as reward-model quality context, not as proof-system fidelity. Fig.~\ref{appx-fig:pairwise-agreement-trend} plots the pairwise service-fidelity trend across model sizes and context lengths, showing that proof-aware agreement with the baseline path stays high for short and medium contexts and declines only at the longest context length under quantization drift.

\section{No-Proof Baseline Latency}
\label{appx-app:additional-latency}

The main text reports optimized per-request artifact generation; Appendix~\ref{appx-app:historical-latency} retains the pre-optimization full-P4 breakdown. Table~\ref{appx-tab:baseline-latency-revised} gives the historical steady-state no-proof baseline for the E01 minimum settings. This baseline is the trace-aware quantized serving path exposed by the BN254 prototype, not a vanilla HuggingFace forward-pass benchmark.

\begin{table}[!t]
\centering
\scriptsize
\caption{Benchmark preference accuracy reported only as reward-model context. Evidence: \textsc{D}.}
\label{appx-tab:benchmark-context}
\resizebox{\columnwidth}{!}{%
\begin{tabular}{llllll}
\toprule
Model & Seq & Baseline Benchmark Acc. & Proof Benchmark Acc. & Difference & Comment \\
\midrule
1B & 128 & 63.50\% & 63.50\% & 0 pp & context only \\
1B & 256 & 56\% & 54.50\% & -1.50 pp & context only \\
1B & 512 & 61\% & 62.50\% & 1.50 pp & context only \\
1B & 1024 & 68.50\% & 70\% & 1.50 pp & context only \\
3B & 128 & 59.50\% & 60.50\% & 1 pp & context only \\
3B & 256 & 57\% & 57\% & 0 pp & context only \\
3B & 512 & 66.50\% & 66\% & -0.50 pp & context only \\
3B & 1024 & 75.50\% & 77.50\% & 2 pp & context only \\
8B & 128 & 59.50\% & 61\% & 1.50 pp & context only \\
8B & 256 & 67\% & 66.50\% & -0.50 pp & context only \\
8B & 512 & 78.50\% & 76.50\% & -2 pp & context only \\
8B & 1024 & 87\% & 83\% & -4 pp & context only \\
\bottomrule
\end{tabular}%
}
\end{table}

\begin{table}[!t]
\centering
\scriptsize
\caption{Historical no-proof baseline before optimization; not a matched baseline for the optimized witness path. Revised steady-state no-proof baseline latency for the E01 minimum settings. Evidence: \textsc{D}. Values are means over non-warmup runs; P95 uses nearest-rank over 20 runs. \brev{Here \emph{Raw RM} is the steady-state no-proof BN254 serving forward time averaged over 20 non-warmup baseline-only runs; it excludes trace recording, proof generation, P4 binding, and export, and is the reference latency for the \emph{S3 Raw RM} column observed inside the with-proof runs.}}
\label{appx-tab:baseline-latency-revised}
\resizebox{\columnwidth}{!}{%
\begin{tabular}{llrrrllrr}
\toprule
Model & Seq & Runs & Tokenize & Quant/Encode & Raw RM & Trace Off Baseline & Baseline Mean & Baseline P95 \\
\midrule
1B & 128 & 20 & 0.0035 & 0.0027 & 26.16 & excluded & 26.18 & 27.81 \\
1B & 1024 & 20 & 0.0036 & 0.0155 & 122.08 & excluded & 122.09 & 125.87 \\
3B & 512 & 20 & 0.0038 & 0.0124 & 186.45 & excluded & 186.46 & 189.92 \\
8B & 128 & 20 & 0.0038 & 0.0050 & 165.45 & excluded & 165.47 & 166.61 \\
8B & 1024 & 20 & 0.0039 & 0.0280 & 832.44 & excluded & 832.47 & 839.22 \\
\bottomrule
\end{tabular}%
}
\end{table}

\section{\texorpdfstring{\brev{Public-Attestation Matrix}}{Public-Attestation Matrix}}
\label{appx-app:additional-public-verification}

The main manuscript visualizes the gas trend for the public wrapper path. Table~\ref{appx-tab:public-verification-full} reports the strict 12-setting \brev{public-attestation} matrix, including payload size, calldata size, wrapper-generation time, and the accept result for each setting.

\begin{table}[!t]
\centering
\scriptsize
\caption{\brev{Strict public-attestation matrix for the implemented wrapper interface, aggregated over three artifacts per setting. Evidence: \textsc{R}. Acceptance records wrapper generation, Solidity verification, gas, proof size, and statement-binding behavior; complete composite verification remains the off-chain auditor path.}}
\label{appx-tab:public-verification-full}
\resizebox{\columnwidth}{!}{%
\begin{tabular}{llrrrrrrr}
\toprule
Seq & Model & Runs & Gas & PayloadB & ProofB & CalldataB & WrapperS & Accept \\
\midrule
128 & 1B & 3 & 217218 & 28580 & 164 & 11573 & 0.3085 & ACCEPT \\
128 & 3B & 3 & 216885 & 49316 & 164 & 19564 & 0.4405 & ACCEPT \\
128 & 8B & 3 & 216881 & 56228 & 164 & 22229 & 0.5254 & ACCEPT \\
256 & 1B & 3 & 217218 & 28580 & 164 & 11573 & 0.2879 & ACCEPT \\
256 & 3B & 3 & 216889 & 49316 & 164 & 19565 & 0.3866 & ACCEPT \\
256 & 8B & 3 & 216893 & 56228 & 164 & 22229 & 0.5087 & ACCEPT \\
512 & 1B & 3 & 217227 & 28580 & 164 & 11573 & 0.3022 & ACCEPT \\
512 & 3B & 3 & 216881 & 49316 & 164 & 19564 & 0.4445 & ACCEPT \\
512 & 8B & 3 & 216877 & 56228 & 164 & 22229 & 0.5269 & ACCEPT \\
1024 & 1B & 3 & 216893 & 28580 & 164 & 11573 & 0.2893 & ACCEPT \\
1024 & 3B & 3 & 216897 & 49316 & 164 & 19565 & 0.4326 & ACCEPT \\
1024 & 8B & 3 & 216897 & 56228 & 164 & 22230 & 0.5240 & ACCEPT \\
\bottomrule
\end{tabular}%
}
\end{table}

\section{Service-Bound Rejection Tests}
\label{appx-app:additional-rejection}

Tables~\ref{appx-tab:request-binding-tests} and~\ref{appx-tab:negative-artifacts} give the detailed rejection evidence behind RQ4 in the main text. They separate request-metadata checks in the artifact-verification harness from broader artifact substitutions and public-tuple replay. The complete-model settings are 1B/128, 3B/512, and 8B/1024; 13 attacks and one valid control per setting give 42 expected outcomes. Export-digest and verifier-ID substitutions are checked off chain, whereas wrapper mismatches are checked under the partial public relation. The request-metadata tests recompute the request identifier and do not imply that the wrapper constrains the complete export or input preprocessing.

\begin{table}[!t]
\centering
\scriptsize
\caption{Request-binding tests on complete artifacts at three representative settings. Metadata substitutions are checked by recomputing request identity; public-field replay also exercises the partial wrapper. PASS denotes the expected outcome, including acceptance of the valid control. Evidence: \textsc{R}.}
\label{appx-tab:request-binding-tests}
\resizebox{\columnwidth}{!}{%
\begin{tabular}{lllllllll}
\toprule
Model & Seq & Valid & Wrong RequestID & Wrong Prompt Digest & Wrong Response Digest & New Nonce Replay & New Policy Replay & Gas Delta \\
\midrule
1B & 128 & PASS & PASS & PASS & PASS & PASS & PASS & 13763 \\
3B & 512 & PASS & PASS & PASS & PASS & PASS & PASS & 13776 \\
8B & 1024 & PASS & PASS & PASS & PASS & PASS & PASS & 13752 \\
\bottomrule
\end{tabular}%
}
\end{table}

\begin{table}[!t]
\centering
\scriptsize
\caption{Negative service-artifact tests for request, model, configuration, verifier, proof-order, digest, and wrapper-statement substitutions. Evidence: \textsc{R}. Each of three settings has 13 attacks and one valid control; PASS denotes the expected outcome. Export-digest and verifier checks are off chain.}
\label{appx-tab:negative-artifacts}
\resizebox{\columnwidth}{!}{%
\begin{tabular}{lllll}
\toprule
Attack & Expected & Observed & Failure Location & Recorded PASS \\
\midrule
valid artifact & ACCEPT & ACCEPT & none & PASS \\
tampered proof & REJECT & REJECT & composite proof & PASS \\
wrong reward & REJECT & REJECT & reward-head chain & PASS \\
wrong model commitment & REJECT & REJECT & model commitment binding & PASS \\
wrong weight opening & REJECT & REJECT & weight opening / weight-task link & PASS \\
wrong configID & REJECT & REJECT & config binding / lookup-table version & PASS \\
wrong verifierID & REJECT & REJECT & verifier binding & PASS \\
wrong requestID & REJECT & REJECT & request binding / replay policy & PASS \\
reordered proof tasks & REJECT & REJECT & op\_order / transcript binding & PASS \\
wrong export digest & REJECT & REJECT & off-chain export digest & PASS \\
stale artifact replay & REJECT & REJECT & request binding / replay policy & PASS \\
wrong public statement & REJECT & REJECT & wrapper relation / model commitment & PASS \\
missing requestID & REJECT & REJECT & statement schema / request binding & PASS \\
mismatched wrapper proof & REJECT & REJECT & Groth16 wrapper & PASS \\
\bottomrule
\end{tabular}%
}
\end{table}

\section{Service-Operation Evidence}
\label{appx-app:additional-service-operation}

This section retains the original service-operation records as historical evidence. The componentized amortization model in Table~\ref{appx-tab:weight-amortization} treats binding generation as reusable preparation and uses measured binding verification for the link/check term. Because opening generation depends on the current request transcript, this split does not establish a valid cached producer cost for the implemented protocol. The batching and queueing results below inherit that assumption and are conditional service scenarios, not measured performance of the optimized implementation.

\begin{table}[!t]
\centering
\scriptsize
\caption{Historical conditional model, not measured cached producer latency; transcript-dependent opening generation must be paid per request. \brev{Componentized P4 amortization from measured timings. Evidence: \textsc{S} built from \textsc{D}.}}
\label{appx-tab:weight-amortization}
\resizebox{\columnwidth}{!}{%
\begin{tabular}{llrrrrrrr}
\toprule
Model & Seq & Runs & Full P4 & \brev{Link/check} & N10 & N50 & N100 & Cache MB \\
\midrule
1B & 128 & 10 & 1152.3 & 9.20 & 123.5 & 32.06 & 20.63 & 11.81 \\
1B & 1024 & 10 & 1015.5 & 4.33 & 105.5 & 24.55 & 14.44 & 12.91 \\
3B & 512 & 10 & 2276.6 & 7.87 & 234.7 & 53.24 & 30.56 & 22.66 \\
8B & 128 & 10 & 3848.6 & 9.14 & 393.1 & 85.93 & 47.53 & 24.67 \\
8B & 1024 & 10 & 3839.0 & 9.12 & 392.1 & 85.72 & 47.42 & 26.89 \\
\bottomrule
\end{tabular}%
}
\end{table}

The batching replay in Table~\ref{appx-tab:batching-throughput} then shows how service windows change request throughput when several requests share checkpoint-level preprocessing assumptions.

\begin{table}[!t]
\centering
\scriptsize
\caption{Historical conditional service-window model using the original amortization assumption; not optimized-producer throughput. Service-window batching replay using measured full-P4 stage rows. Evidence: \textsc{S} built from \textsc{D}. \brev{Values are computed from 20 service-window replays per cell using measured single-request full-P4 stage rows; this is not a true tensor-batch benchmark, as the prototype does not expose a multi-request batch mode.}}
\label{appx-tab:batching-throughput}
\resizebox{\columnwidth}{!}{%
\begin{tabular}{llrrrrrr}
\toprule
Model & Seq & B & Windows & Mean Batch & P95 Req & Req/hour & GPU MB \\
\midrule
1B & 256 & 1 & 20 & 1443.0 & 1448.0 & 2.50 & 19591 \\
1B & 256 & 2 & 20 & 1853.7 & 927.8 & 3.88 & 19591 \\
1B & 256 & 4 & 20 & 2674.9 & 669.2 & 5.38 & 19591 \\
1B & 256 & 8 & 20 & 4317.6 & 539.9 & 6.67 & 19591 \\
1B & 1024 & 1 & 20 & 1565.8 & 1596.1 & 2.30 & 39539 \\
1B & 1024 & 2 & 20 & 2120.4 & 1078.7 & 3.40 & 39539 \\
1B & 1024 & 4 & 20 & 3229.7 & 819.5 & 4.46 & 39539 \\
1B & 1024 & 8 & 20 & 5448.2 & 686.0 & 5.29 & 39539 \\
3B & 512 & 1 & 20 & 3171.3 & 3243.1 & 1.14 & 29031 \\
3B & 512 & 2 & 20 & 4074.0 & 2070.3 & 1.77 & 29031 \\
3B & 512 & 4 & 20 & 5879.3 & 1488.3 & 2.45 & 29031 \\
3B & 512 & 8 & 20 & 9489.8 & 1194.7 & 3.04 & 29031 \\
8B & 512 & 1 & 20 & 5128.1 & 5130.1 & 0.70 & 44119 \\
8B & 512 & 2 & 20 & 6412.2 & 3208.9 & 1.12 & 44119 \\
8B & 512 & 4 & 20 & 8980.4 & 2246.5 & 1.60 & 44119 \\
8B & 512 & 8 & 20 & 14116.8 & 1765.2 & 2.04 & 44119 \\
\bottomrule
\end{tabular}%
}
\end{table}

Auditor-side cost is reported separately in Table~\ref{appx-tab:auditor-workload}, which replays verification capacity from real exported artifacts rather than from a synthetic verifier model.

\begin{table}[!t]
\centering
\scriptsize
\caption{Historical auditor replay before optimization. Auditor-side verification capacity from real artifact replay. Evidence: \textsc{R}. \brev{Capacity is a replay over retained real artifacts (the per-setting artifact count is shown in the Artifacts column); concurrency $C$ scales worker capacity analytically and is not an independently measured multi-worker deployment benchmark.}}
\label{appx-tab:auditor-workload}
\resizebox{\columnwidth}{!}{%
\begin{tabular}{llrrrrrr}
\toprule
Model & Seq & C & Artifacts & Mean Verify & P95 Verify & Artifacts/hour & Memory MB \\
\midrule
1B & 128 & 1 & 41 & 11.97 & 13.71 & 301.1 & 11.8 \\
1B & 128 & 4 & 41 & 11.97 & 13.71 & 1204.5 & 47.3 \\
1B & 128 & 16 & 41 & 11.97 & 13.71 & 4818.2 & 189.0 \\
1B & 1024 & 1 & 10 & 5.80 & 5.89 & 621.2 & 12.9 \\
1B & 1024 & 4 & 10 & 5.80 & 5.89 & 2484.9 & 51.7 \\
1B & 1024 & 16 & 10 & 5.80 & 5.89 & 9939.5 & 206.7 \\
3B & 512 & 1 & 10 & 10.38 & 10.54 & 346.7 & 22.7 \\
3B & 512 & 4 & 10 & 10.38 & 10.54 & 1386.9 & 90.7 \\
3B & 512 & 16 & 10 & 10.38 & 10.54 & 5547.8 & 362.7 \\
8B & 128 & 1 & 10 & 11.85 & 12.02 & 303.8 & 24.7 \\
8B & 128 & 4 & 10 & 11.85 & 12.02 & 1215.1 & 98.7 \\
8B & 128 & 16 & 10 & 11.85 & 12.02 & 4860.6 & 394.8 \\
8B & 1024 & 1 & 10 & 12.14 & 12.27 & 296.5 & 26.9 \\
8B & 1024 & 4 & 10 & 12.14 & 12.27 & 1185.8 & 107.6 \\
8B & 1024 & 16 & 10 & 12.14 & 12.27 & 4743.3 & 430.4 \\
\bottomrule
\end{tabular}%
}
\end{table}

Finally, Table~\ref{appx-tab:service-latency-throughput} uses the measured stage latencies in a trace-driven queueing replay to illustrate conditional cache-friendly scenarios under the original amortization assumptions; these are not validated deployment regimes for the optimized producer.

\begin{table}[!t]
\centering
\scriptsize
\caption{Historical conditional simulation; cache-based results are not SLA measurements of the optimized implementation. Trace-driven queueing results under delayed-audit arrivals. Evidence: \textsc{S}.}
\label{appx-tab:service-latency-throughput}
\resizebox{\columnwidth}{!}{%
\begin{tabular}{llrrrrrrl}
\toprule
Model & Seq & B & $\lambda$ & Cache hit & Off-chain P95 & Public P95 & Throughput & SLA met \\
\midrule
1B & 256 & 2 & 0.1 & 0.95 & 22.21h & 22.21h & 13.87 & miss \\
1B & 256 & 2 & 0.5 & 0.95 & 4.52h & 4.52h & 13.87 & both \\
1B & 256 & 2 & 1.0 & 0.95 & 1.92h & 1.92h & 13.87 & both \\
3B & 512 & 2 & 0.1 & 0.95 & 29.09h & 29.09h & 5.11 & miss \\
3B & 512 & 2 & 0.5 & 0.95 & 4.63h & 4.63h & 5.11 & both \\
3B & 512 & 2 & 1.0 & 0.95 & 3.03h & 3.03h & 5.11 & both \\
8B & 512 & 2 & 0.1 & 0.95 & 29.25h & 29.25h & 2.74 & miss \\
8B & 512 & 2 & 0.5 & 0.95 & 4.79h & 4.79h & 2.74 & both \\
8B & 512 & 2 & 1.0 & 0.95 & 3.18h & 3.18h & 2.74 & both \\
\bottomrule
\end{tabular}%
}
\end{table}

\section{Design Ablation}
\label{appx-app:additional-ablation}

This historical ablation separates mechanisms that change proving cost from bindings that define the accepted service output. Its timing values precede the acceleration work. The no-matmul row is analytical, and cache-based P4 rows inherit the conditional amortization assumption explained in Appendix~\ref{appx-app:additional-service-operation}. Table~\ref{appx-tab:component-ablation} records the retained component rows used by RQ6.

\begin{table}[!t]
\centering
\scriptsize
\caption{Historical before-optimization rows; cache variants are conditional models. Component ablation from real stage measurements and specified removed-component variants. Evidence: \textsc{S} built from \textsc{D}. \brev{Rows are the mean over three seq=512 settings (1B/512, 3B/512, 8B/512). \texttt{Full VRMaaS} P4 is the $N{=}10$ value under the componentized P4 model; the no-cache variant replaces it with measured full P4 per request, and cached/link-check does not mean reusable Gemini/KZG opening generation. The no-matmul-aggregation row is an analytical $1.35\times$ P2 ablation rather than a separately executed prover run. These means over seq=512 settings are not directly comparable to the single-setting E01 latency rows.}}
\label{appx-tab:component-ablation}
\resizebox{\columnwidth}{!}{%
\begin{tabular}{lrrrrrrl}
\toprule
Variant & P2 & P3 & P4 & Payload & Gas & Fidelity & Security / Service Property \\
\midrule
Full VRMaaS & 628.7 & 1.90 & 245.0 & 44708 & 216995 & 96.7\% / R2 0.962 & complete binding + public attestation \\
No matmul aggregation & 848.8 & 1.90 & 245.0 & 44708 & 216995 & 96.7\% / R2 0.962 & aggregation performance optimization removed \\
No weight-binding cache & 628.7 & 1.90 & 2385.6 & 44708 & 216995 & 96.7\% / R2 0.962 & \brev{componentized P4 amortization benefit} \\
Off-chain only & 628.7 & 1.90 & 245.0 & 44708 & 0 & 96.7\% / R2 0.962 & public settlement / on-chain attestation \\
\bottomrule
\end{tabular}%
}
\end{table}

\section{Generic Baseline Attack Outcomes}
\label{appx-app:additional-generic-baseline}

RQ6 also needs a baseline that isolates accepted-statement semantics rather than prover throughput. Table~\ref{appx-tab:generic-attack-outcomes} gives the minimal runnable GenericProof+Wrapper comparison: attacks accepted by GenericProof+Wrapper are not claims that generic proof systems cannot be extended, but evidence that the reduced baseline statement does not bind those service fields.

\begin{table}[!t]
\centering
\scriptsize
\caption{Minimal runnable generic baseline attack outcomes. Evidence: \textsc{R}. ACCEPT in the generic column indicates that the service field is not bound by that baseline statement.}
\label{appx-tab:generic-attack-outcomes}
\resizebox{\columnwidth}{!}{%
\begin{tabular}{lllll}
\toprule
Model & Attack & GenericProof+Wrapper & VRMaaS-reduced & Service implication \\
\midrule
tiny-rm-l1 & tampered proof & REJECT / BLOCKED & REJECT / PASS & proof bytes are checked \\
tiny-rm-l1 & wrong reward & REJECT / BLOCKED & REJECT / PASS & output is checked \\
tiny-rm-l1 & wrong requestID & ACCEPT / ACCEPTED BUT UNBOUND & REJECT / PASS & VRMaaS binds artifact to request \\
tiny-rm-l1 & wrong configID & ACCEPT / ACCEPTED BUT UNBOUND & REJECT / PASS & VRMaaS binds numerical configuration \\
tiny-rm-l1 & wrong model commitment & ACCEPT / ACCEPTED BUT UNBOUND & REJECT / PASS & VRMaaS binds checkpoint commitment \\
tiny-rm-l1 & reordered proof tasks & ACCEPT / ACCEPTED BUT UNBOUND & REJECT / PASS & VRMaaS binds deterministic proof trace \\
tiny-rm-l1 & wrong verifierID & ACCEPT / ACCEPTED BUT UNBOUND & REJECT / PASS & VRMaaS binds verifier implementation \\
tiny-rm-l1 & stale request replay & ACCEPT / ACCEPTED BUT UNBOUND & REJECT / PASS & VRMaaS binds request freshness \\
tiny-transformer-l2 & tampered proof & REJECT / BLOCKED & REJECT / PASS & proof bytes are checked \\
tiny-transformer-l2 & wrong reward & REJECT / BLOCKED & REJECT / PASS & output is checked \\
tiny-transformer-l2 & wrong requestID & ACCEPT / ACCEPTED BUT UNBOUND & REJECT / PASS & VRMaaS binds artifact to request \\
tiny-transformer-l2 & wrong configID & ACCEPT / ACCEPTED BUT UNBOUND & REJECT / PASS & VRMaaS binds numerical configuration \\
tiny-transformer-l2 & wrong model commitment & ACCEPT / ACCEPTED BUT UNBOUND & REJECT / PASS & VRMaaS binds checkpoint commitment \\
tiny-transformer-l2 & reordered proof tasks & ACCEPT / ACCEPTED BUT UNBOUND & REJECT / PASS & VRMaaS binds deterministic proof trace \\
tiny-transformer-l2 & wrong verifierID & ACCEPT / ACCEPTED BUT UNBOUND & REJECT / PASS & VRMaaS binds verifier implementation \\
tiny-transformer-l2 & stale request replay & ACCEPT / ACCEPTED BUT UNBOUND & REJECT / PASS & VRMaaS binds request freshness \\
\bottomrule
\end{tabular}%
}
\end{table}

\section{Configuration Diagnostics}
\label{appx-app:additional-config}

The configuration diagnostic supports the main text's point that $\mathsf{configID}$ is a deployment-facing admission decision, not cosmetic metadata. Table~\ref{appx-tab:quant-config-tradeoff} reports the current-configuration quantization-drift record used for that decision.

\begin{table}[!t]
\centering
\scriptsize
\caption{Current-configuration quantization drift and deployment recommendation. Evidence: \textsc{D}. \brev{Drift is computed once over 200 cases per setting under the single currently registered configuration; it is a single-config diagnostic, not a repeated or multi-configuration variance study.}}
\label{appx-tab:quant-config-tradeoff}
\resizebox{\columnwidth}{!}{%
\begin{tabular}{llrrrrrl}
\toprule
Model & Seq & Config & Mean Abs Err & P95 Abs Err & Pairwise Agree & P2 Cost & Recommendation \\
\midrule
1B & 512 & current config & 0.568 & 1.514 & 96.5\% & 199.5 & register current config \\
1B & 1024 & current config & 0.909 & 2.512 & 93.5\% & 251.2 & register with warning \\
3B & 512 & current config & 0.742 & 2.249 & 97.5\% & 415.3 & register with warning \\
3B & 1024 & current config & 1.730 & 4.908 & 95.0\% & 556.5 & register with warning \\
8B & 512 & current config & 0.939 & 3.268 & 96.0\% & 695.5 & register with warning \\
8B & 1024 & current config & 2.326 & 7.332 & 94.0\% & 955.2 & flag; finer config recommended \\
\bottomrule
\end{tabular}%
}
\end{table}

\section{Case-Level Fidelity Diagnostics}
\label{appx-app:case-level-fidelity}

The aggregate fidelity metrics in the main text measure whether proof-aware serving preserves the observable behavior of the baseline RM service. Table~\ref{appx-tab:case-level-fidelity} gives compact examples from the original RewardBench2 diagnostic case log. Each row abbreviates a prompt-response comparison and reports which side was preferred by the baseline RM path and by the proof-aware \vrmaas{} path. These examples are not evidence that the RM's semantic preference is correct; they only illustrate decision-level agreement between the unproved service output and the proof-carrying service artifact.

\begin{table}[!t]
  \centering
  \caption{Compact case-level fidelity examples from the RewardBench2 diagnostic log. ``Chosen'' and ``Rejected'' denote the two candidate responses in the benchmark pair; the table reports which candidate each serving path ranked higher. \brev{C7--C8 are the two large-quantization-error flips from the 8B/1024 diagnostic log.}}
  \label{appx-tab:case-level-fidelity}
  \scriptsize
  \setlength{\tabcolsep}{2pt}
  \begin{tabular}{p{0.08\columnwidth}p{0.28\columnwidth}p{0.12\columnwidth}p{0.12\columnwidth}p{0.28\columnwidth}}
    \toprule
    \textbf{Case} & \textbf{Prompt summary} & \textbf{Baseline} & \textbf{\vrmaas{}} & \textbf{Diagnostic reading} \\
    \midrule
    C1 & Wikipedia-style article about Toonattik & Chosen & Chosen & Agreement on a factual encyclopedic generation task; the proof-aware path preserves the same pairwise direction. \\
    C2 & CBBC schedule for May 2024 using shows aired in 2023 & Rejected & Rejected & Agreement when both serving paths rank the benchmark's rejected candidate higher, reminding that fidelity is distinct from semantic correctness. \\
    C3 & Past Simple versus Present Perfect lesson exercise & Chosen & Chosen & Agreement on an educational grammar prompt with short explanatory responses. \\
    C4 & Performance of a 2010 computer across 2001--2019 & Chosen & Chosen & Agreement on a structured long-form comparison request. \\
    C5 & Most used programming language in 2024 & Rejected & Rejected & Agreement on a temporally sensitive information request, again measuring service equivalence rather than factuality. \\
    C6 & Story outline and scene image prompts for a periodic-table themed story & Chosen & Chosen & Agreement on a creative generation task with multi-part output constraints. \\
    \brev{C7} & \brev{Evidence about poor people in Mesopotamian empires} & \brev{Rejected} & \brev{Chosen} & \brev{Flip at 8B/1024 (\texttt{case\_id=177}, \texttt{large\_quant\_\allowbreak{}error}); margin changes from -1.50 (29.00 vs 30.50) to +2.20 (37.82 vs 35.62).} \\
    \brev{C8} & \brev{Most recent events known before training cutoff} & \brev{Chosen} & \brev{Rejected} & \brev{Flip at 8B/1024 (\texttt{case\_id=135}, \texttt{large\_quant\_\allowbreak{}error}); margin changes from +1.625 (25.75 vs 24.125) to -1.191 (26.973 vs 28.165).} \\
    \bottomrule
  \end{tabular}
\end{table}

The examples show why the paper treats fidelity as a service property. \vrmaas{} is not intended to improve or reinterpret the RM's preference judgment. Instead, when a downstream auditor receives an accepted artifact, the service-level claim is that the public reward or ranking decision is tied to the same configured RM computation that the provider claims to have run. Cases where both paths agree with each other but disagree with the benchmark label, such as C2 and C5, are therefore not failures of proof-carrying verification; they expose ordinary RM-quality limitations that remain outside the cryptographic service-output guarantee. \brev{For 8B/1024, the diagnostic sweep records 188/200 agreement cases; the 12 flips consist of 10 small-margin cases and the two large-quantization-error cases C7--C8.}

\section{Security and Service-Integrity Mapping}
\label{appx-app:security-mapping}

Table~\ref{appx-tab:threat-guarantee} maps service risks to the checks, formal statements, and tests supporting the implemented interfaces. Complete artifact acceptance is an off-chain decision; public-tuple replay has the narrower guarantee of the partial wrapper theorem. Appendix~\ref{appx-app:extended-attestation} separately specifies the conditional export-binding extension.

\begin{table*}[t]
  \centering
  \scriptsize
  \setlength{\tabcolsep}{2pt}
  \caption{Service risks, the \vrmaas{} checks that address them, and where each is established formally and tested empirically.}
  \label{appx-tab:threat-guarantee}
  \begin{tabular}{@{}p{0.17\textwidth}p{0.55\textwidth}p{0.24\textwidth}@{}}
    \toprule
    \textbf{Risk} & \textbf{Check and accepted-artifact guarantee} & \textbf{Formal / test} \\
    \midrule
    Wrong score & $\Pi_{\mathrm{comp}}$ binds $r_{\mathrm{pub}}$ to the last registered reward-head task; acceptance implies a valid hidden execution. & Lemma~\ref{lem:rm-soundness} / Table~\ref{appx-tab:negative-artifacts} \\
    Model swap & $\mathcal{C}_W$, openings $\mathcal{B}_W$, and the weight-task link authenticate every proof-relevant weight operand. & Lemma~\ref{lem:rm-soundness}, \eqref{eq:weight-task-link} / Table~\ref{appx-tab:negative-artifacts} \\
    Config/verifier swap & $\mathsf{configID}$ fixes tables, scales, ranges, and export rule; $\mathsf{verifierID}$ fixes the verifier code. & Lemma~\ref{lem:rm-soundness}; Appendix~\ref{appx-app:fiat-shamir} / Table~\ref{appx-tab:negative-artifacts} \\
    Request replay & The off-chain verifier checks request metadata against the expected invocation and absorbs the identifier into the transcript; the wrapper separately checks the supplied tuple. & \eqref{eq:request-id}; Appendix~\ref{appx-app:fiat-shamir} / Table~\ref{appx-tab:negative-artifacts} \\
    Proof reorder & $\mathsf{op\_order}$ is absorbed into the transcript; batch membership is recorded. & Lemma~\ref{lem:rm-soundness} / Table~\ref{appx-tab:negative-artifacts} \\
    Public tuple replay & The wrapper's public-field equalities bind one proof to one supplied public tuple; export binding remains an auditor check. & Thm.~\ref{thm:public-attestation-soundness} / Table~\ref{appx-tab:negative-artifacts} \\
    Public leakage & The contract sees only $\mathrm{Stmt}^{\mathrm{proto}}_{\mathrm{pub}}$ and $\pi_G$; inputs, weights, and openings stay witness-side. & Lemma~\ref{lem:public-privacy} \\
    \bottomrule
  \end{tabular}
\end{table*}

\section{Related-Work Comparison}
\label{appx-app:related-work}

Table~\ref{appx-tab:related-comparison} summarizes the verifier-facing statement semantics that distinguish \vrmaas{} from adjacent lines of work. The comparison is not a claim that \vrmaas{} is universally faster than generic ZKML or zkLLM systems. It records which fields are treated as first-class accepted statement fields rather than as deployment metadata.

\begin{table}[!t]
  \centering
  \scriptsize
  \setlength{\tabcolsep}{1.5pt}
  \caption{Verifier-facing statement semantics of \vrmaas{} and adjacent approaches in their original problem formulations. ``First-class'' means that the field is part of the exported object or accepted verifier statement, not merely external metadata; the table is not a claim that generic proof systems cannot be extended with such fields.}
  \label{appx-tab:related-comparison}
  \begin{tabular}{@{}p{0.36\columnwidth}cccccc@{}}
    \toprule
    \textbf{Approach / refs} &
    \textbf{Req.} &
    \textbf{Wt.} &
    \textbf{Cfg.} &
    \textbf{Vrf.} &
    \textbf{Order} &
    \textbf{Export} \\
    \midrule
    RM evaluation\newline \cite{malik2025rewardbench2advancingreward,liu2025skywork} &
    -- & -- & -- & -- & -- & -- \\
    Verifiable rewards\newline \cite{liu2025trust,cobbe2021training} &
    \(\circ\) & -- & \(\circ\) & -- & -- & -- \\
    Generic ZKML / MLaaS\newline \cite{parno2016pinocchio,ghodsi2017safetynets,zhao2021veriml,huang2022zkmlaas} &
    \(\circ\) & \(\circ\) & \(\circ\) & -- & -- & \(\circ\) \\
    zkLLM systems\newline \cite{sun2024zkllm,qu2025zkgpt,liao2026verilora} &
    \(\circ\) & \(\circ\) & \(\circ\) & -- & \(\circ\) & -- \\
    Trustworthy services\newline \cite{huhns2005service,papazoglou2007service,froelicher2020drynx} &
    \(\checkmark\) & -- & \(\circ\) & \(\circ\) & -- & \(\checkmark\) \\
    \textbf{\vrmaas{} (this work)} &
    \(\checkmark\) & \(\checkmark\) & \(\checkmark\) &
    \(\checkmark\) & \(\checkmark\) & \(\checkmark\) \\
    \bottomrule
  \end{tabular}

  \vspace{0.3em}
  \footnotesize
  Req.: request identity; Wt.: model-weight commitment and openings; Cfg.: numerical configuration and lookup tables; Vrf.: verifier implementation; Order: proof-task order; Export: verifier-facing artifact or public statement. \(\circ\) denotes partial support or external metadata. The \vrmaas{} row describes the complete off-chain acceptance interface; the public wrapper exposes only the tuple in Equation~\eqref{eq:public-statement}.
\end{table}

Reward-model evaluation and verifiable-reward methods focus on semantic reliability, benchmark behavior, or externally checkable task outcomes. Verifiable ML and ZKML systems prove neural-network computation on untrusted infrastructure and can protect private inputs, model parameters, or intermediate states. Trustworthy service and MLaaS work addresses cross-organization service composition, outsourced computation, integrity assurance, and payment. \vrmaas{} is complementary to these threads: it uses proof-system machinery to make the RM service output itself portable, request-bound, model-bound, configuration-bound, and consumable by auditors or settlement layers.

\section{Historical Producer Latency Before Optimization}
\label{appx-app:historical-latency}
\begin{figure}[t]
\centering
\includegraphics[width=\columnwidth]{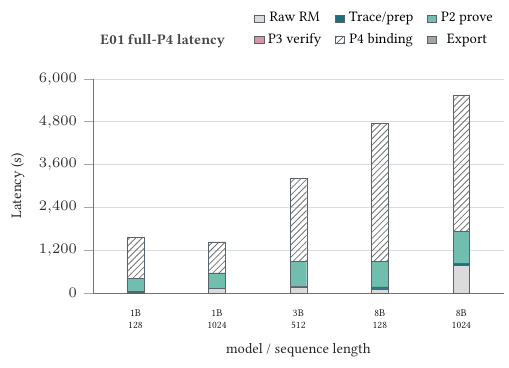}
\caption{Historical stage breakdown before optimization, retained from the original manuscript. The optimized per-request measurements are reported in Fig.~\ref{fig:latency-breakdown}.}
\label{fig:historical-latency}
\end{figure}

Table~\ref{tab:withproof-latency-revised} retains the original artifact-producing latency records, and Fig.~\ref{fig:historical-latency} retains their stage visualization. These are before-optimization measurements. They are preserved to document the engineering starting point, not presented as costs of the optimized implementation. P2 denotes computation proving, P3 producer-side verification, and P4 weight binding. The before/after comparison in the main text uses matching model/sequence settings for the binding stage while stating the differences in proof coverage and execution conditions.

\begin{table}[t]
\centering
\scriptsize
\caption{Historical measurements before optimization. Proof-aware serving latency and stage breakdown for representative artifact-producing runs. Evidence: \textsc{D}. The 1B/128 row averages 41 completed non-warmup repetitions; the remaining rows are retained full-P4 artifact-producing runs.}
\label{tab:withproof-latency-revised}
\resizebox{\columnwidth}{!}{%
\begin{tabular}{llrrrrrrrr}
\toprule
Model & Seq & Runs & \brev{S3 Raw RM} & Trace/Preprocess & P2 & P3 & P4 & Export & Total \\
\midrule
1B & 128 & 41 & 26.59 & 15.09 & 363.95 & 2.47 & 1156.53 & 0.44 & 1565.06 \\
1B & 1024 & 1 & 123.16 & 8.03 & 432.60 & 1.21 & 867.60 & 0.16 & 1432.76 \\
3B & 512 & 1 & 170.88 & 23.52 & 701.12 & 2.12 & 2319.00 & 0.35 & 3216.98 \\
8B & 128 & 1 & 114.34 & 57.67 & 711.78 & 2.24 & 3861.40 & 0.74 & 4748.16 \\
8B & 1024 & 1 & 778.27 & 58.65 & 899.55 & 2.53 & 3802.00 & 0.75 & 5541.75 \\
\bottomrule
\end{tabular}%
}
\end{table}

\section{Export-Binding Public Attestation as an Extension}
\label{appx-app:extended-attestation}
The main text specifies the implemented partial wrapper and the relation measured by the public-attestation experiments. This appendix retains the stronger export-binding design as a conditional extension. Implementing it requires constraining the canonical export digest, request-metadata hash, and decoded field equalities inside the wrapper; these checks are not supplied by the measured public tuple. Its costs are not represented by the reported 217K gas. Even this extension retains a compact attestation predicate: complete computation verification must still come from an off-chain audit of the same digest-bound export unless the full verifier is also proved inside the circuit.

\subsection{Exported Public Statement}
The extended relation derives the public statement from the exported package $\mathcal{E}$. The off-chain prototype canonically encodes the package and computes a BN254-native Poseidon2 digest:
Its canonical digest is defined in Equation~\eqref{eq:export-digest}.
The digest $\mathsf{H}_{E}$ is a compact public handle for the exported package. It is not the Fiat--Shamir transcript for the inner proof checks; those challenges are derived during composite-proof verification from the ordered proof-task transcript described above.
The exported package contains more than the public reward: it includes the request identifier, ordered proof-task encoding, compact proof data, the model-weight commitment, and the private weight-binding material consumed by the exported verifier. In the framework-level relation, the selected public fields below are submitted to the contract; the full proof transcript and weight-binding material remain off chain. The public statement is
\begin{equation}
\mathrm{Stmt}^{\mathrm{ext}}_{\mathrm{pub}}
=
\left(
\mathsf{requestID},
r_{\mathrm{pub}},
\mathsf{H}_{E},
\mathcal{C}_{W},
\mathsf{verifierID},
\mathsf{configID}
\right).
\label{appx-ext-eq:public-statement}
\end{equation}
Here, $\mathsf{requestID}$ binds the statement to the service job, $r_{\mathrm{pub}}$ is the published reward, $\mathsf{H}_{E}$ binds the statement to the exported package, $\mathcal{C}_{W}$ binds the intended model weights, and the identifiers bind the verifier logic and numerical configuration. A contract implementing this extension accepts $\mathcal{C}_{W}$ only when the commitment is registered by the model publisher or otherwise authenticated for the intended RM service. Because $\mathsf{requestID}$ is included in both $\mathcal{E}$ and $\mathrm{Stmt}^{\mathrm{ext}}_{\mathrm{pub}}$, an accepted reward statement cannot be silently grafted onto another request without breaking the export digest or the wrapper relation.

\subsection{Extended Wrapper Relation}
The on-chain path does not replay the RM forward pass or expose the inner proof transcript. \brev{The complete exported verifier, denoted $V_{\vrmaas}(\mathcal{E})$, is run by the off-chain auditor to check the composite proof, proof-task order, configuration identifiers, and weight-binding material.} Instead, a Groth16 wrapper~\cite{groth2016} \brev{can implement a compact attestation predicate $\mathsf{Attest}_{\mathrm{wrap}}$ over the exported package and the public statement.} The wrapper relation is
\begin{equation}
\begin{aligned}
\mathcal{R}^{\mathrm{ext}}_{\mathrm{wrap}}:\quad
\exists \mathcal{E}, w_{\mathrm{wrap}}\ \mathrm{s.t.}\quad
&\mathsf{H}_{E}=\mathsf{Digest}_{254}(\mathcal{E}),\\
&\brev{\mathsf{Attest}_{\mathrm{wrap}}(\mathcal{E};w_{\mathrm{wrap}})=1,}\\
&\yrev{\mathsf{Hash}(\mathsf{ReqMeta}(\mathcal{E}))=\mathsf{requestID},}\\
&\mathsf{RequestID}(\mathcal{E})=\mathsf{requestID},\\
&\mathsf{Reward}(\mathcal{E})=r_{\mathrm{pub}},\\
&\mathsf{ModelCommit}(\mathcal{E})=\mathcal{C}_{W},\\
&\mathsf{VerifierID}(\mathcal{E})=\mathsf{verifierID},\\
&\mathsf{ConfigID}(\mathcal{E})=\mathsf{configID}.
\end{aligned}
\label{appx-ext-eq:wrapper-relation}
\end{equation}
The witness of the wrapper circuit is $(\mathcal{E},w_{\mathrm{wrap}})$, where $w_{\mathrm{wrap}}$ contains decoding witnesses and intermediate values needed to arithmetize \brev{the compact attestation predicate}. \brev{Equivalently, the circuit view is $C_{\mathrm{wrap}}(\mathrm{Stmt}^{\mathrm{ext}}_{\mathrm{pub}};\mathcal{E},\mathrm{aux})=1$, where $\mathrm{aux}$ contains arithmetization intermediates and does not change the deterministic verifier semantics of $V_{\vrmaas}(\mathcal{E})$.} A circuit implementing this extension must enforce the checks in \eqref{appx-ext-eq:wrapper-relation}: the private package hashes to the public digest, \brev{the selected fold/sumcheck and reward-consistency checks accepted by $\mathsf{Attest}_{\mathrm{wrap}}$ hold inside $\mathcal{E}$,} the request identifier, reward, and model commitment decoded from $\mathcal{E}$ match the public statement, and the verifier/configuration identifiers match the declared deployment configuration. For such a circuit, the prover generates
\begin{equation}
\pi_G
\leftarrow
\mathsf{Groth16.Prove}
\left(
C_{\mathrm{wrap}},
\mathrm{Stmt}^{\mathrm{ext}}_{\mathrm{pub}};
\mathcal{E},w_{\mathrm{wrap}}
\right),
\label{appx-ext-eq:groth16-wrapper-proof}
\end{equation}
where $C_{\mathrm{wrap}}$ implements \eqref{appx-ext-eq:wrapper-relation}. A contract implementing this extension receives only $\mathrm{Stmt}^{\mathrm{ext}}_{\mathrm{pub}}$ and $\pi_G$. A typical acceptance rule is
\begin{equation}
\begin{aligned}
\mathsf{Accept}^{\mathrm{ext}}_{\mathrm{chain}}
=&\ \mathsf{Registered}(\mathcal{C}_{W})\\
&\wedge
\mathsf{Verify}_{\mathrm{Groth16}}
(\mathsf{vk}_{G},\mathrm{Stmt}^{\mathrm{ext}}_{\mathrm{pub}},\pi_{G}).
\end{aligned}
\label{appx-ext-eq:chain-accept}
\end{equation}
After acceptance of this extended relation, the reward becomes a contract-facing output under its stated checks. The wrapper contract can return an acceptance bit, return the accepted reward value, or emit an event containing the accepted statement digest for later audit. These interfaces expose the result of \eqref{appx-ext-eq:chain-accept}; they do not add another proof step and they do not reveal the private wrapper witness. Thus, public acceptance means that the statement is backed by an exported package satisfying the compact attestation relation and bound to the declared request identity, model commitment, verifier implementation, and numerical configuration. \brev{The full composite-verifier result remains an off-chain auditor check, and the public statement gives downstream contracts a compact settlement handle derived from the same export.}

\begin{theorem}[Conditional export-binding attestation]
\label{appx-ext-thm:public-attestation-soundness}
If the contract accepts \eqref{appx-ext-eq:chain-accept}, then, except with negligible probability, $\mathrm{Stmt}^{\mathrm{ext}}_{\mathrm{pub}}$ is bound to a private exported package satisfying the compact wrapper-attestation relation under the authenticated model commitment, request identity, verifier implementation, and numerical configuration. \brev{When an off-chain auditor also accepts $V_{\vrmaas}(\mathcal{E})=1$, the same export links this public statement to the complete RM computation proof.}
\end{theorem}

\begin{proof}[Proof sketch]
By Groth16 soundness, an accepting $\pi_G$ implies a private exported package $\mathcal{E}$ and wrapper witness $w_{\mathrm{wrap}}$ satisfying \eqref{appx-ext-eq:wrapper-relation}. The wrapper relation \brev{checks $\mathsf{Attest}_{\mathrm{wrap}}(\mathcal{E};w_{\mathrm{wrap}})=1$,} recomputes $\mathsf{H}_E=\mathsf{Digest}_{254}(\mathcal{E})$, and checks that the request identifier, reward, model commitment, verifier identifier, and configuration identifier decoded from $\mathcal{E}$ match $\mathrm{Stmt}^{\mathrm{ext}}_{\mathrm{pub}}$. \brev{Since \eqref{appx-ext-eq:wrapper-relation} also checks $\mathsf{Hash}(\mathsf{ReqMeta}(\mathcal{E}))=\mathsf{requestID}$, the accepted package is bound to its recorded request metadata; this check alone does not prove that tokenization and embedding lookup produced the supplied initial hidden states.} The digest and field-equality checks prevent statement rebinding or cross-request grafting except by digest collision, proof forgery, \yrev{input-digest collision,} or commitment-binding failure. \brev{If an auditor additionally runs and accepts the complete off-chain verifier, Lemma~\ref{lem:rm-soundness} supplies the RM-computation soundness for that same export.}
\end{proof}

\bibliographystyle{IEEEtran}
\bibliography{main}

\end{document}